\documentclass[twocolumn,showpacs,preprintnumbers,amsmath,amssymb,showkeys]{revtex4}
\usepackage{graphicx}
\usepackage{dcolumn}
\usepackage{bm}

\begin{document}

\title{Intraband memory function and memory-function conductivity formula in doped graphene}
 \author{I. Kup\v{c}i\'{c}}     
 \address{
   Department of Physics, Faculty of Science, University of Zagreb, 
   P.O. Box 331,  HR-10002 Zagreb,  Croatia}

\begin{abstract} 
The generalized self-consistent field method is used to describe intraband relaxation processes in 
a general multiband electronic system with presumably weak residual electron-electron interactions.
	The resulting memory-function conductivity formula is shown to have the same structure 
as the result of a more accurate approach based on the quantum kinetic equation.
	The results are applied to heavily doped and lightly doped graphene.
	It is shown that the scattering of conduction electron by phonons leads to the redistribution 
of the intraband conductivity spectral weight over a wide frequency range, 
however, in a way consistent with the partial transverse conductivity sum rule.
	The present form of the intraband memory function is found to describe correctly the scattering 
by quantum fluctuations of the lattice, at variance with the semiclassical Boltzmann transport equations,
where this scattering channel is absent.
	This is shown to be of fundamental importance in quantitative understanding of 
the reflectivity data measured in lightly doped graphene as well as in different low-dimensional 
strongly correlated electronic systems, such as the cuprate superconductors.
\end{abstract}
%
%
\maketitle

\section{Introduction}
In condensed matter physics, important information can be obtained about
interactions in the electronic subsystem by analyzing
relaxation processes associated with the scattering of conduction electrons by static disorder, by phonons, and by other electrons.
      One of the central questions regarding
the relaxation processes is to explain temperature and retardation effects in simple physical terms 
by using simple enough self-consistent kinetic equations.
	The memory function is the common name for the ${\bf k}$- and $\omega$-dependent relaxation function 
in such self-consistent kinetic equations \cite{Kubo95,Forster75,Gotze72,Allen71,Kupcic14,Kupcic15}.
	The relaxation rate is its imaginary part at zero frequency.
	The memory function is usually introduced to describe intraband relaxation processes in the dynamical conductivity tensor,
in the Raman response functions, as well as in different transport coefficients.
	It is well known that even in weakly interacting systems the explanation of experimental observations
requires a unified diagrammatic representation for
the so-called self-energy contributions to the response function in question and the related vertex corrections
\cite{Mahan90,Ziman88,Kupcic15}.
	Moreover, it is easily seen that more complicated electronic system is longer is the list of requirements that the response functions 
and the relaxation functions in question must satisfy.
	The causality principle, the law of conservation of energy, and the charge continuity equation 
are all of fundamental importance in understanding the relaxation processes.
	Consequently, they play an important role in analyzing measured transport coefficients and measured reflectivity 
and Raman scattering spectra by means of such self-consistent kinetic equations.

The {\it stosszahl ansatz} in Boltzmann transport equations represents the simplest way to explain qualitatively the temperature 
dependence of the intraband relaxation rate \cite{Kubo95,Pines89,Ziman72,Abrikosov88}.
	The part of the relaxation rate associated with the scattering by phonons is proportional to the Bose-Einstain distribution 
function and the $1/2$ term associated with corresponding quantum fluctuations of the lattice is missing.
	As a result, the Boltzmann transport equations have serious deficiencies in describing
the retardation effects, in particular those associated with the scattering by optical phonons and by other high-energy boson modes.

The generalized Drude formula is the primary tool for investigating retardation effects.
	It is usually assumed to be a model independent method of analyzing measured reflectivity and Raman scattering spectra
in terms of the $\omega$-dependent memory function $M_{\alpha} (\omega)$ \cite{Uchida91,Degiorgi91,Schwartz98,Mirzaei13,Opel00,Basov11}.
	However, in most cases of general interest the extraction of $M_{\alpha} (\omega)$ from experimental data depends on details 
in the boson mediated electron-electron interactions, on general properties of the crystal potential, as well as on the very nature of the local field effects.
	Consequently, such an analysis is usually incomplete and often inadequate.
	Therefore, to study the interband conductivity, or the excitations across the 
charge-density-wave (CDW), spin-density-wave (SDW) or superconducting Bardeen-Cooper-Schrieffer (BCS) gap or pseudogap, we need 
general enough self-consistent kinetic equations and much more sophisticated procedures for solving
these equations than that usually used to derive the generalized Drude formula.

Lightly doped graphene is an important weakly interacting two-band system in which the threshold energy for interband electron-hole 
excitations is of the order of optical phonon energies, the optical phonon energies are quite large, and the intraband and interband
contributions to the dynamical conductivity tensor are expected to be decoupled from each other \cite{Li08,Castro09}.
	The structure of the dynamical conductivity is similar to that of typical CDW/SDW pseudogaped systems, and,
consequently, lightly doped graphene is a convenient model system for reexamining different open questions 
regarding electrodynamics of conduction electrons in such multiband electronic systems.

In this paper, we use the generalized self-consistent field method [usually called the generalized random-phase approximation (RPA)] to derive the 
memory-function conductivity formula for the intraband conductivity and to determine the structure of the intraband 
memory function in heavily doped and lightly doped graphene.
	The results are compared to both the results of the common variational method for the dc conductivity \cite{Ziman72}
and to the results of a more accurate approach based on the quantum kinetic equation \cite{Kupcic13,Kupcic15}.
	It is shown that the scattering by phonons leads to the redistribution of the conductivity spectral weight 
over a wide energy range in a way consistent with the partial transverse conductivity sum rule.
	The intraband memory function has the same structure as that obtained by means of the quantum kinetic equation.

The paper is organized as follows.
	In Sec. II, we briefly describe all elements in the total Hamiltonian for conduction electrons in a general multiband case.
To make the reading of the paper easier, we give in Secs.~III and IV an overview of both the macroscopic identity relations 
among the exact elements of the real-time RPA irreducible $4 \times 4$ response tensor and the microscopic version of the same identity relations.
	The partial effective mass theorem and the related transverse conductivity sum rule are shown to play an essential role in determining
the proper structure of the memory-function conductivity formula.
	This transverse conductivity sum rule can also be useful in reexamining gauge invariance of the conductivity formula obtained by means 
of the common current-current approach \cite{Ando02,Peres08,Carbotte10} or by different charge-charge approaches \cite{Hwang07,Despoja13}.
	In Secs. IV and V, we discuss general properties of the generalized self-consistent RPA equations 
and the quantum transport equations.
	These two equations are used in Sec.~VI to derive the intraband memory-function conductivity formula 
and the leading contributions to the intraband memory function.
	The relation between the memory-function conductivity formula and the generalized Drude conductivity formula is briefly discussed in Sec.~VII.
	In Sec.~VIII, the numerical results for the real and imaginary parts of the intraband memory function are presented for heavily doped graphene
for typical values of the model parameters.
	In Sec.~IX, we consider the two-band conductivity in lightly doped graphene.
	In this section, the emphasis is on the appropriate parametrization of the low-energy intraband conductivity tensor and on the connection 
between the effective generalized Drude formula obtained in this way and the aforementioned partial transverse conductivity sum rule.
	Section X contains concluding remarks.

\section{Model Hamiltonian}
In electronic systems with multiple bands in the vicinity of the Fermi level, 
conduction electrons are described by the Hamiltonian \cite{Kupcic15}
\begin{eqnarray}
&& \hspace{-2mm}
H = H^{\rm el}_0 + H^{\rm ph}_0 + H'_{1a} +  H'_{1b} + H'_2 + H^{\rm ext}.
\label{eq1}
\end{eqnarray} 
	The bare electronic contribution
\begin{eqnarray}
&& \hspace{-10mm}
H_0^{\rm el} = 
\sum_{L {\bf k} \sigma} [\varepsilon_L^0 ({\bf k})+ \mu] c^\dagger_{L {\bf k} \sigma} c_{L {\bf k} \sigma}
\label{eq2}
\end{eqnarray}
represents noninteracting electrons in such a multiband case.
	Here, $\varepsilon_L^0 ({\bf k})$ is the bare electron dispersion measured with respect to the chemical potential 
$\mu$ in the band labeled by the band index $L$.
	$H^{\rm ph}_0$ is the bare phonon Hamiltonian
\begin{eqnarray}
&& \hspace{-10mm}
H^{\rm ph}_0 = \sum_{\lambda {\bf q}'} \frac{1}{2M_\lambda} \big[  p^{\dagger}_{\lambda {\bf q}'} p_{\lambda {\bf q}'} + \big( 
M_\lambda \omega_{\lambda {\bf q}'} \big)^2 u^{\dagger}_{\lambda {\bf q}'} u_{\lambda {\bf q}'}\big]
\label{eq3}
 \end{eqnarray} 
given in terms of the phonon field $u_{\lambda {\bf q}'}$, and the conjugate field $p_{\lambda {\bf q}'}$,
$\omega_{\lambda {\bf q}'}$ is the bare phonon frequency, $\lambda$ is the phonon branch index, 
and $M_\lambda$ is the corresponding effective ion mass.

The electron-phonon coupling Hamiltonian can be shown in the following way
\begin{eqnarray}
&& \hspace{-5mm}
H'_{1a} = \sum_{\lambda LL'} \sum_{{\bf k} {\bf q}' \sigma} \frac{G_\lambda^{L'L} ({\bf k}_+,{\bf k})}{
\sqrt{N}} \big(b_{\lambda {\bf q}'} + b^\dagger_{\lambda -{\bf q}'} \big)
c^{\dagger}_{L'{\bf k}+{\bf q}' \sigma} c_{L{\bf k} \sigma},  
\nonumber \\
\label{eq4}
\end{eqnarray} 
where $u_{\lambda {\bf q}'} = \sqrt{(\hbar/2M_\lambda \omega_{\lambda {\bf q}'})} (b_{\lambda {\bf q}'} + b^\dagger_{\lambda -{\bf q}'} )$
and ${\bf k}_+ = {\bf k} + {\bf q}'$.
	This expression includes the scattering by acoustic and optical phonons.
	On the other hand, the scattering by static disorder is given by 
\begin{eqnarray}
&& \hspace{-5mm}
H'_{1b} =    \sum_{LL'} \sum_{{\bf k} {\bf q}'\sigma} 
V^{L'L}({\bf k}_+,{\bf k})
c^{\dagger}_{L'{\bf k}+{\bf q}' \sigma} c_{L{\bf k} \sigma}.
\label{eq5}
\end{eqnarray} 
	 Finally, the electron-electron interaction Hamiltonian
\begin{eqnarray}
&& \hspace{-2mm}
H'_2 =  \frac{1}{2V} \sum_{LL'L_1L_1'} \sum_{{\bf k}{\bf k}'{\bf q}} \sum_{\sigma \sigma'} 
\varphi_{\sigma \sigma'}^{L'L_1L_1'L} ({\bf q}) 
\nonumber \\
&& \hspace{7mm} \times 
c^{\dagger}_{L'{\bf k}+{\bf q} \sigma} c^{\dagger}_{L_1{\bf k}'\sigma'}
c_{L_1'{\bf k}'+{\bf q}\sigma'} c_{L{\bf k} \sigma}
\label{eq6}
\end{eqnarray} 
describes all nonretarded electron-electron interactions.

The coupling between conduction electrons and external electromagnetic fields is obtained by 
the gauge-invariant tight-binding minimal substitution
\cite{Schrieffer64,Kupcic13}.
	The result is $H^{\rm ext} = H^{\rm ext}_1 + H^{\rm ext}_2$, where
\begin{eqnarray}
 && \hspace{-12mm}
H^{\rm ext}_1  =  \sum_{{\bf q}}  V^{\rm ext} ({\bf q}) \hat \rho (-{\bf q}) 
-\frac{1}{c} \sum_{{\bf q} \alpha} A_{\alpha}^{\rm ext} ({\bf q})
\hat J_{\alpha} (-{\bf q}),
\nonumber \\
&& \hspace{-12mm}
H^{\rm ext}_2 =  \frac{e^2}{2mc^2} \sum_{{\bf q} {\bf q}'\alpha \beta} 
A_{\alpha}^{\rm ext} ({\bf q}-{\bf q}')A_{\beta}^{\rm ext} ({\bf q}') \hat  \gamma_{\alpha \beta}(-{\bf q};2)
\label{eq7}
 \end{eqnarray} 
($\alpha, \beta = x,y,z$ in a general three-dimensional case).
	Here, $V^{\rm ext} ({\bf q}, \omega)$ and ${\bf A}^{\rm ext}({\bf q}, \omega)$ are, respectively, 
the Fourier transforms of the external scalar and vector potentials, while
the corresponding screened potentials are labeled by $V^{\rm tot} ({\bf r}, t)$ and ${\bf A}^{\rm tot}({\bf r}, t)$. 
	The total charge density operator in the coupling Hamiltonian (\ref{eq7}) is
\begin{eqnarray}
\hat \rho ({\bf q}) \equiv \hat J_{0} ({\bf q})
= \sum_{LL'}\sum_{{\bf k}\sigma} e q^{LL'} ({\bf k}, {\bf k}_+)
c^\dagger_{L{\bf k} \sigma} c_{L'{\bf k}+{\bf q} \sigma}.
\label{eq8}
\end{eqnarray}
	The structures of the corresponding current density operator  $\hat J_{\alpha} ({\bf q})$
and the bare diamagnetic density operator $\hat \gamma_{\alpha \beta}({\bf q};2)$ are similar.
	Finally, $eq^{LL'} ({\bf k}, {\bf k}_+) \equiv J_0^{LL'} ({\bf k}, {\bf k}_+)$, $J_\alpha^{LL'} ({\bf k}, {\bf k}_+)$, and 
$\gamma_{\alpha \beta}^{LL'} ({\bf k},{\bf k}_+;2)$ are the bare vertex functions in question.
	Hereafter, the dispersions $\varepsilon_L^0 ({\bf k})$ and all these vertex functions are taken as known functions
(for doped graphen see, for example, Ref.~\cite{Kupcic13}).

\section{Kubo formula for conductivity tensor}
Electrodynamic properties of multiband electronic systems
are naturally described in terms of the screened dynamical conductivity tensor 
\begin{eqnarray}
&& \hspace{-10mm}
\widetilde \sigma_{\alpha \beta} ({\bf q}, \omega) = 
\beta \int_0^\infty d t\, e^{{\it i} \omega t} \frac{1}{V} 
\big< \hat J_\beta (-{\bf q},0); \hat J_\alpha ({\bf q},t) \big> .
\label{eq9}
\end{eqnarray}
	This relation is known as the Kubo formula for conductivity \cite{Kubo95}.
	The conductivity tensor $\sigma_{\alpha \beta} ({\bf q}, \omega)$ is simply the RPA irreducible part of 
$\widetilde \sigma_{\alpha \beta} ({\bf q}, \omega)$.
	In those multiband electronic systems in which Lorentz local field effects are absent
(the two-band model for $\pi$ electrons in graphene from Sec.~VIII being an example), the result is
\begin{eqnarray}
&& \hspace{-5mm}
\sigma_{\alpha \beta} ({\bf q}, \omega) = 
\beta \int_0^\infty d t\, e^{{\it i} \omega t} \frac{1}{V} 
\big< \hat J_\beta (-{\bf q},0); \hat J_\alpha ({\bf q},t) \big>_{\rm irred}.
\nonumber \\
\label{eq10}
\end{eqnarray}
	This form of $\sigma_{\alpha \beta} ({\bf q}, \omega)$ holds in the single-band case as well, because
there are no local field effects in this case.

One usually uses the definition relation (\ref{eq10}) and the two basic relations from macroscopic electrodynamics,
\begin{eqnarray}
&& \hspace{-10mm}
{\bf E}({\bf r}, t) = - \frac{\partial V^{\rm tot} ({\bf r}, t) }{\partial {\bf r}}
-\frac{1}{c} \frac{\partial {\bf A}^{\rm tot}({\bf r}, t)}{\partial t},
\label{eq11}
\\
&& \hspace{-10mm}
\nabla \cdot {\bf J} ({\bf r}, t) + \frac{\partial \rho ({\bf r}, t)}{\partial t} = 0,
\label{eq12}
\end{eqnarray}
to show $\sigma_{\alpha \beta} ({\bf q}, \omega)$ in terms of the elements of the
real-time RPA irreducible $4 \times 4$ response tensor 
\begin{eqnarray}
&& \hspace{-11mm}
V \pi_{\mu \nu}({\bf q},t) = -\frac{{\it i}}{\hbar} \theta(t) 
\big< \big[ \hat J_\mu ({\bf q},t), \hat J_\nu (-{\bf q},0) \big] \big>_{\rm irred}
\label{eq13}
\end{eqnarray}
($\mu, \nu = 0, x,y$ in graphene,
and $\mu, \nu = 0, x,y,z$ in a general three-dimensional case)
and the real-time current-dipole correlation function $\pi_{\alpha \tilde \beta} ({\bf q}, t)$,
rather than in terms of the correlation functions 
$\big< \hat J_\beta (-{\bf q},0); \hat J_\alpha ({\bf q},t) \big>_{\rm irred}$.
	The result is \cite{Kubo95,Kupcic14}
\begin{eqnarray}
&& \hspace{-4mm}
\pi_{00} ({\bf q}, \omega) 
= \frac{1}{\omega} \sum_{\beta} \pi_{0 \beta} ({\bf q}, \omega) q_\beta
=  \frac{1}{{\it i}\omega} \sum_{\alpha \beta} q_\alpha \sigma_{\alpha \beta} ({\bf q}, \omega) q_\beta,
\nonumber \\
\label{eq14} \\
&& \hspace{-5mm}
{\it i} \pi_{\alpha 0} ({\bf q}, \omega) = 
\frac{{\it i}}{\omega} \sum_{\beta} \big[ \pi_{\alpha \beta} ({\bf q}, \omega)-\pi_{\alpha \beta} ({\bf q})\big] q_\beta
\nonumber \\
&& \hspace{11mm}
= \sum_{\beta}  \sigma_{\alpha \beta} ({\bf q}, \omega) q_\beta, 
\label{eq15}
\\
&& \hspace{-5mm}
\sigma_{\alpha \beta} ({\bf q}, \omega) =  \pi_{\alpha \tilde \beta} ({\bf q}, \omega).
\label{eq16}
\end{eqnarray}
	Here, we have introduced the notation 
$\hat J_{\tilde \alpha} ({\bf q}) = - \hat P_{\alpha} ({\bf q})$, where $\hat P_{\alpha} ({\bf q})$ 
is the dipole density operator and $P^{LL'}_{\alpha} ({\bf k}, {\bf k}_+)$ 
is the corresponding dipole vertex function \cite{Kupcic13}.
	Equation (\ref{eq14}), for example, shows that the conductivity tensor $\sigma_{\alpha \beta} ({\bf q}, \omega)$, divided by ${\it i} \omega$,  
is nothing but the second-order coefficient in the Taylor expansion of the charge-charge correlation function
$\pi_{00} ({\bf q}, \omega)$ with respect to $q_\alpha$.

In the simplest case with longitudinal electromagnetic fields, where ${\bf q} = q_\alpha \hat e_\alpha$,
the conductivity tensor from Eqs.~(\ref{eq14})$-$(\ref{eq16}) becomes
\begin{eqnarray}
&& \hspace{-5mm}
\sigma_{\alpha \alpha} ({\bf q}, \omega)  
= \frac{{\it i}}{q_\alpha}  \pi_{\alpha 0} ({\bf q}, \omega) = \frac{{\it i} \omega}{q_\alpha^2}\pi_{00} ({\bf q}, \omega) 
=  \pi_{\alpha \tilde \alpha} ({\bf q}, \omega), 
\nonumber \\
&& \hspace{-5mm}
\sigma_{\alpha \alpha} ({\bf q}, \omega)   = 
\frac{{\it i}}{\omega} \big[ \pi_{\alpha \alpha} ({\bf q}, \omega)-\pi_{\alpha \alpha} ({\bf q})\big]. 
\label{eq17}
\end{eqnarray}
	Since $\sigma_{\alpha \alpha} ({\bf q}, \omega) $ is a non-singular function of ${\bf q}$ and $\omega$ for all
${\bf q}$ and $\omega$, the elements of the $4 \times 4$ response tensor are expected to have the properties
\begin{eqnarray}
&& \hspace{-5mm}
\pi_{00} ({\bf q}, \omega) \propto q_\alpha^2, \hspace{3mm} \pi_{0\alpha} ({\bf q}, \omega) \propto q_\alpha,
\hspace{3mm}
{\rm Im} \{\pi_{\alpha \alpha} ({\bf q}, \omega) \} \propto \omega.
\nonumber \\
\label{eq18}
\end{eqnarray}
	These relations are the usual starting point for
hydrodynamic formulation of electrodynamics of conduction electrons \cite{Forster75,Gotze72,Vollhardt80}.
	They prove useful in systematic microscopic studies
of $\sigma_{\alpha \alpha} ({\bf q}, \omega)$ as well \cite{Kupcic15,Kupcic16}.

\subsection{Partial transverse conductivity sum rule}
For transverse electromagnetic fields polarized along the $\alpha$ axis, we can write
\begin{eqnarray}
&& \hspace{-10mm}
\sigma_{\alpha \alpha} ({\bf q}, \omega)  
=  \pi_{\alpha \tilde \alpha} ({\bf q}, \omega) 
= \frac{{\it i}}{\omega} \big[ \pi_{\alpha \alpha} ({\bf q}, \omega)-\pi_{\alpha \alpha} ({\bf q})\big]. 
\label{eq19}
\end{eqnarray}
	After performing the Kramers-Kronig analysis \cite{Kubo95}, the transverse conductivity sum rule becomes
a function of the static current-current correlation function $\pi_{\alpha \alpha} ({\bf q})$, 
\begin{eqnarray}
&& \hspace{-10mm}
4 \int_{-\infty} ^\infty d \omega \, {\rm Re} \{ \sigma_{\alpha \alpha}({\bf q},\omega) \}
= - 4 \pi \pi_{\alpha \alpha} ({\bf q}).
\label{eq20}
\end{eqnarray}
	From the multiband version of the Ward identity relation \cite{Kupcic14}, it follows that
\begin{eqnarray}
&& \hspace{-5mm}
- 4 \pi \pi_{\alpha \alpha} ({\bf q}) = \frac{4 \pi e^2}{m} n^{\rm tot}_{\alpha \alpha} ({\bf q}).
\label{eq21}
\end{eqnarray}
	The quantity 
\begin{eqnarray}
&& \hspace{-5mm} n^{\rm tot}_{\alpha \alpha}({\bf q}) = \sum_{LL'}
\frac{1}{V} \sum_{{\bf k} \sigma} \frac{m}{e^2}
\frac{|J_\alpha^{LL'} ({\bf k},{\bf k}_+)|^2}{\varepsilon^0_{L'L} ({\bf k}_+,{\bf k})} 
[n_L({\bf k}) - n_{L'}({\bf k}_+) ]
\nonumber \\
&& \hspace{7mm} 
= n_{\alpha \alpha}^{\rm intra} ({\bf q}) + n_{\alpha \alpha}^{\rm inter} ({\bf q})
\label{eq22}
\end{eqnarray}
in Eq.~(\ref{eq21}) is the total effective number of charge carriers, which comprises the intraband contribution 
$n_{\alpha \alpha}^{\rm intra} ({\bf q})$ ($L=L'$) and the interband contribution $n_{\alpha \alpha}^{\rm inter} ({\bf q})$ ($L \neq L'$)
\cite{Kupcic16}.

For long wavelengths, the effective number $n_{\alpha \alpha}^{\rm intra} ({\bf q})$ 
can be rewritten in the alternative form, in terms of the 
dimensionless reciprocal effective mass tensor $\gamma^{LL}_{\alpha \alpha}({\bf k}) = (m/\hbar^2) \partial^2 \varepsilon_L^0({\bf k})/\partial k_\alpha^2$.
	In this limit, the total effective number $n^{\rm tot}_{\alpha \alpha} ({\bf q}\approx {\bf 0})$ becomes 
\begin{eqnarray}
&& \hspace{-5mm} n^{\rm tot}_{\alpha \alpha} ({\bf q}\approx {\bf 0})
= \frac{1}{V} \sum_{L{\bf k} \sigma} \gamma_{\alpha \alpha}^{LL} ({\bf k} ;2) n_L({\bf k}),
\label{eq23}
\end{eqnarray}
where \cite{Kupcic07,Kupcic13,Kupcic14}
\begin{eqnarray}
&& \hspace{-10mm}
\gamma^{LL}_{\alpha \alpha}({\bf k};2) = \gamma^{LL}_{\alpha \alpha}({\bf k}) 
+ \frac{m}{e^2} \sum_{L'(\neq L)} \frac{2 |J_\alpha^{L L'}({\bf k})|^2 }{
\varepsilon^0_{L'L}({\bf k},{\bf k})}.
\label{eq24}
 \end{eqnarray} 
	Therefore, the sum rule (\ref{eq20}) is in accordance with the partial effective mass theorem (\ref{eq24}) linking
the bare diamagnetic vertex 
$\gamma_{\alpha \alpha}^{LL} ({\bf k},{\bf k}_+;2) \approx \gamma_{\alpha \alpha}^{LL} ({\bf k};2)$ with
the reciprocal effective mass tensor $\gamma^{LL}_{\alpha \alpha}({\bf k})$ and the interband current vertices 
$J_\alpha^{L L'}({\bf k},{\bf k}_+) \approx J_\alpha^{L L'}({\bf k})$.

In Eqs.~(\ref{eq22}) and (\ref{eq23}), $n_L({\bf k})$ is the momentum distribution function defined by
\begin{eqnarray}
&& \hspace{-10mm}
n_L({\bf k}) = \frac{1}{\beta \hbar} \sum_{{\it i}\omega_n} {\cal G}_L ({\bf k}, {\it i} \omega_n)
=  \int_{-\infty}^\infty \frac{d \varepsilon}{2 \pi} \, {\cal A}_L ({\bf k}, \varepsilon) f(\varepsilon).
\label{eq25}
\end{eqnarray}
	Here, $f(\varepsilon)$ is the Fermi-Dirac distribution function,
${\cal G}_L ({\bf k}, {\it i} \omega_n)$ is the single-electron Green's function,	
and ${\cal A}_L ({\bf k}, \varepsilon)$ is the corresponding spectral function.
	${\cal G}_L ({\bf k}, {\it i} \omega_n)$ is the Matsubara Fourier transform of 
${\cal G}_{L} ({\bf k}, \tau) = - \langle T_\tau [c_{L{\bf k} \sigma} (\tau) c^\dagger_{L{\bf k} \sigma} (0) ] \rangle$.

The sum rule (\ref{eq20}) must not be confused with the usual form of the transverse conductivity sum rule, 
which can be found in the literature \cite{Pines89,Mahan90}.
	The latter represents the generalization of Eq.~(\ref{eq20}) to the case with infinite number of valence bands.
	In this case, the effective number $n^{\rm tot}_{\alpha \alpha} ({\bf q})$ reduces to the nominal concentration
of conduction electrons $n$ [$\gamma^{LL}_{\alpha \alpha}({\bf k};2)=1$ for the conduction band, in this case].

The partial version of the sum rule holds for any electronic system with finite number of valence bands 
which is decoupled from the rest of the band structure.
	Evidently the partial transverse conductivity sum rule is much more useful in investigations of
tight-binding systems with a few bands  [where $n^{\rm tot}_{\alpha \alpha} ({\bf q})$ is usually very different from $n$]
than its common textbook version.
	In this case, the left-hand side and the right-hand side of Eq.~(\ref{eq20}) can be calculated independently
providing the direct test of the conductivity formula used in the calculations.

\begin{figure}
   \centerline{\includegraphics[width=20pc]{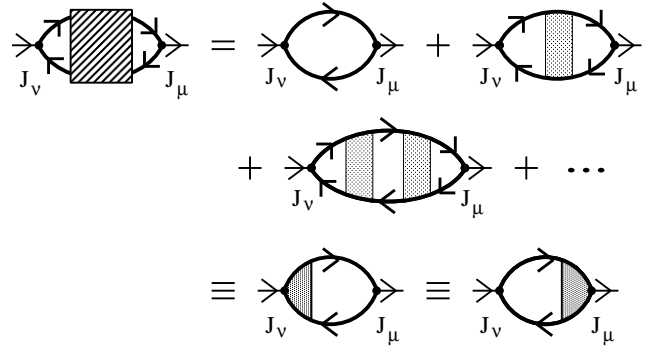}}
   \caption{The Bethe-Salpeter expression for the $4 \times 4$ current-current correlation function
   $\pi_{\mu \nu} ({\bf q}, {\it i} \nu_n)$ \cite{Schrieffer64,Ziman88,Kupcic15}.
   The bold solid lines represent the exact single-electron Green's functions and the shaded rectangle represents the exact RPA
   irreducible four-point interaction.
   }
  \end{figure}
  
\section{Theoretical approaches}
\subsection{Bethe-Salpeter equations}
In realistic electronic systems with multiple bands, the microscopic structure of the conductivity tensor 
$\sigma_{\alpha \beta} ({\bf q}, \omega)$ is usually determined by using the Matsubara finite-temperature 
formalism \cite{Abrikosov75,Fetter71,Ziman88,Mahan90}. 
	In this approach, the correlation functions $\pi_{\mu \nu} ({\bf q},\omega)$ from Eqs.~(\ref{eq14})$-$(\ref{eq16}) are obtained 
by analytical continuation of $\pi_{\mu \nu} ({\bf q},{\it i} \nu_n)$ (${\it i} \nu_n \rightarrow \omega + {\it i} \eta$),
where $\pi_{\mu \nu} ({\bf q},{\it i} \nu_n)$ is the Matsubara Fourier transform of
\begin{eqnarray}
&& \hspace{-10mm}
\pi_{\mu \nu} ({\bf q},\tau) 
= - \frac{1}{\hbar V} \big<  T_\tau \big[\hat J_\mu ({\bf q}, \tau) \hat J_{\nu} (-{\bf q}, 0) \big] \big>_{\rm irred}.
\label{eq26}
\end{eqnarray}
	According to Fig.~1, $\pi_{\mu \nu} ({\bf q},{\it i} \nu_n)$ is shown in terms of the exact single-electron 
Green's function ${\cal G}_L ({\bf k}, {\it i}\omega_n)$ and the exact RPA irreducible four-point interaction
$U^{L_1L'L_1'L} ({\bf k}_+,{\bf k}',{\bf k}'_+, {\bf k}, {\it i} \omega_{n+}, {\it i} \omega_{m},
{\it i} \omega_{m+}, {\it i} \omega_n)$.
	The single-electron Green's function ${\cal G}_L ({\bf k}, {\it i}\omega_n)$ satisfies the Dyson equation, and 
the RPA irreducible four-point interaction the corresponding Bethe-Salpeter equation \cite{Abrikosov75,Fetter71,Ziman88,Mahan90}.
	For many purposes, it is helpful to 
rewrite this Bethe-Salpeter expression for $\pi_{\mu \nu} ({\bf q},{\it i} \nu_n)$ in terms of 
${\cal G}_L ({\bf k}, {\it i}\omega_n)$ and the exact renormalized vertex function $\Gamma_\nu^{L'L}({\bf k}_+,{\bf k},{\it i}\omega_{n+}, {\it i}\omega_{n})$.
	The Bethe-Salpeter equation for $\Gamma_\nu^{L'L}({\bf k}_+,{\bf k},{\it i}\omega_{n+}, {\it i}\omega_{n})$ is closely related to 
that for the RPA irreducible four-point interaction.
	Finally, it is also possible to
show $\pi_{\mu \nu} ({\bf q},{\it i} \nu_n)$ as a function of ${\cal G}_L ({\bf k}, {\it i}\omega_n)$ and 
$\Phi_{\nu}^{LL'} ({\bf k},{\bf k}_+,{\it i}\omega_n,{\it i}\omega_{n+})$, the quantity  which is usually 
called the auxiliary electron-hole propagator \cite{Kupcic13,Kupcic15,Kupcic16}, the three-point electron-hole propagator, or the three-point susceptibility 
\cite{Bergeron11}.

As long as these building blocks of $\pi_{\mu \nu} ({\bf q},{\it i} \nu_n)$ are exact, all Kubo-Ward relations from the previous section are exactly fulfilled.
	This means that, in this case, the correlation functions $\pi_{\mu \nu} ({\bf q},{\it i} \nu_n)$ have a form which is gauge invariant by definition,
and the charge continuity equation is exactly satisfied.
	However, any approximation used to determine the structures of ${\cal G}_L ({\bf k}, {\it i}\omega_n)$ and 
$\Phi_{\nu}^{LL'} ({\bf k},{\bf k}_+,{\it i}\omega_n,{\it i}\omega_{n+})$ leads  to some extent to the violation of the charge continuity equation.
	As a consequence, we are usually forced to take
care of the charge continuity equation explicitly when solving the Dyson and Bethe-Salpeter equations.

\subsection{Generalized self-consistent RPA equations}
In weakly interacting systems, we can also use the alternative approach which represents an obvious generalization
of the common self-consistent RPA equation.
	In this approach, we consider the Heisenberg equation for the density operator $c^\dagger_{L{\bf k} \sigma} c_{L'{\bf k}+{\bf q} \sigma}$
\cite{Kupcic14,Platzman73},
\begin{eqnarray}
&& \hspace{-5mm}
{\it i} \hbar \frac{\partial }{\partial t} c^\dagger_{L{\bf k} \sigma} c_{L'{\bf k}+{\bf q} \sigma} = 
\big[c^\dagger_{L{\bf k} \sigma} c_{L'{\bf k}+{\bf q} \sigma},H \big].
\label{eq27}
\end{eqnarray}
	In the general case, the Hamiltonian $H$ is given by Eq.~(\ref{eq1}).
	Therefore, we can use this approach to study the scattering of conduction electrons by static disorder, by phonons, as well as by other electrons.
	For example, for the relaxation processes associated with the scattering by phonons, a straightforward calculation leads to 
\begin{eqnarray}
&& \hspace{-10mm}
[ \hbar \omega + \varepsilon_{LL'}^0({\bf k},{\bf k}_+)  + {\it i} \eta ]
c^\dagger_{L{\bf k} \sigma} c_{L'{\bf k}+{\bf q} \sigma} 
\nonumber \\
&& \hspace{5mm}
= - \big( \hat n_{L{\bf k} \sigma} - \hat n_{L'{\bf k}+{\bf q} \sigma}  \big) P^{L'L}_\alpha ({\bf k}_+,{\bf k})E_\alpha ({\bf q},\omega)
\nonumber \\
&& \hspace{8mm}
+ [c^\dagger_{L{\bf k} \sigma} c_{L'{\bf k}+{\bf q} \sigma},H_{1a}']
\label{eq28}
\end{eqnarray}
[$\varepsilon_{LL'}^0({\bf k},{\bf k}_+)=\varepsilon_{L}^0({\bf k})-\varepsilon_{L'}^0({\bf k}_+)$].
	Here, $E_\alpha ({\bf r},t)$ is again the macroscopic electric field, the $P^{L'L}_\alpha ({\bf k}_+,{\bf k})$ are the intraband 
and interband dipole vertex functions, and $\hat n_{L{\bf k} \sigma} = c^\dagger_{L{\bf k} \sigma}c_{L{\bf k} \sigma}$.
	To obtain the self-consistent structure of these equations, we have to determine the right-hand side expressions in the equations
\begin{eqnarray}
&& \hspace{-10mm}
{\it i} \hbar \frac{\partial }{\partial t} \big\{ (b_{\lambda {\bf k}+{\bf q}-{\bf k}'} + b^\dagger_{\lambda {\bf k}'-{\bf k}- {\bf q}}) 
c^\dagger_{L{\bf k} \sigma} c_{L'{\bf k}' \sigma} \big\} 
\nonumber \\
&& \hspace{5mm}
= 
\big[(b_{\lambda {\bf k}+{\bf q}-{\bf k}'} + b^\dagger_{\lambda {\bf k}-{\bf k}- {\bf q}}) c^\dagger_{L{\bf k} \sigma} c_{L'{\bf k}' \sigma},H \big],
\nonumber \\
&& \hspace{-10mm}
{\it i} \hbar \frac{\partial }{\partial t} \big\{ (b_{\lambda {\bf k}'-{\bf k}} + b^\dagger_{\lambda {\bf k}-{\bf k}'}) 
c^\dagger_{L{\bf k}' \sigma} c_{L'{\bf k}+{\bf q}\sigma} \big\} 
\nonumber \\
&& \hspace{5mm}
= 
\big[(b_{\lambda {\bf k}'-{\bf k}} + b^\dagger_{\lambda {\bf k}-{\bf k}'}) c^\dagger_{L{\bf k}' \sigma} c_{L'{\bf k}+{\bf q}\sigma},H \big],
\label{eq29}
\end{eqnarray}
and retain only the contributions proportional either to $c^\dagger_{L{\bf k} \sigma} c_{L'{\bf k}+{\bf q} \sigma}$ or to 
$c^\dagger_{L{\bf k}' \sigma} c_{L'{\bf k}'+{\bf q} \sigma}$.
	The former contributions will be referred to as the self-energy contributions and the latter ones as the vertex corrections.
	When the electron does not change the band when it is scattered by phonons
and $G_\lambda^{LL}({\bf k},{\bf k}') \approx G_\lambda({\bf k},{\bf k}')$, then
the result is the self-consistent equation 
for the induced density $\langle c^\dagger_{L{\bf k} \sigma} c_{L'{\bf k}+{\bf q} \sigma} \rangle_\omega$ of the form
\begin{eqnarray}
&& \hspace{-5mm}
[\hbar \omega + \varepsilon_{LL'}^0({\bf k},{\bf k}_+)  + {\it i} \eta ]
\langle c^\dagger_{L{\bf k} \sigma} c_{L'{\bf k}+{\bf q} \sigma} \rangle_\omega
\nonumber \\
&& \hspace{0mm}
= [ n_{L'}({\bf k}_+) - n_{L}({\bf k})] P^{L'L}_\alpha ({\bf k}_+,{\bf k})E_\alpha ({\bf q},\omega)
\nonumber \\
&& \hspace{3mm}
+ \lambda^2 \sum_{\lambda {\bf k}'}\frac{|G_\lambda({\bf k},{\bf k}')|^2}{N}\big[ {\cal S}_{LL'}({\bf k}, {\bf k}',\omega) 
\langle c^\dagger_{L{\bf k} \sigma} c_{L'{\bf k}+{\bf q} \sigma} \rangle_\omega
\nonumber \\
&& \hspace{3mm}
- {\cal S}_{LL'}({\bf k}', {\bf k},\omega) 
\langle c^\dagger_{L{\bf k}' \sigma} c_{L'{\bf k}'+{\bf q} \sigma} \rangle_\omega \big].
\label{eq30}
\end{eqnarray}
	Here,
\begin{eqnarray}
&& \hspace{-5mm}
{\cal S}_{LL'}({\bf k}, {\bf k}',\omega) =
\sum_{s = \pm 1} \frac{f^b (\omega_{\lambda  {\bf k}-{\bf k}'}) 
+ f(s \varepsilon_{L'}^0({\bf k}'))}{
\hbar \omega + {\it i} \eta + \varepsilon_{LL'}^0({\bf k},{\bf k}') + s \hbar \omega_{\lambda {\bf k}-{\bf k}'}}
\nonumber \\
&& \hspace{10mm}
+ \sum_{s = \pm 1} \frac{f^b (\omega_{\lambda  {\bf k}-{\bf k}'}) 
+ f(-s \varepsilon_{L}^0({\bf k}'))}{
\hbar \omega + {\it i} \eta + \varepsilon_{LL'}^0({\bf k}',{\bf k}) + s \hbar \omega_{\lambda {\bf k}-{\bf k}'}}
\label{eq31}
\end{eqnarray}
is a useful abbreviation.

It must be recalled that $\langle c^\dagger_{L{\bf k} \sigma} c_{L'{\bf k} +{\bf q} \sigma} \rangle_\omega  \equiv \delta n^{LL'} ({\bf k}, {\bf q}, \omega)$
is the nonequilibrium part of the nonequilibrium distribution function in question $n^{LL'} ({\bf k}, {\bf q}, \omega)$ \cite{Kupcic14,Platzman73}.
	Therefore, the induced current density can be shown in terms of the current-dipole correlation function $\pi_{\alpha \tilde \alpha}({\bf q}, \omega)$
in the following way
\begin{eqnarray}
&& \hspace{-10mm}
J_\alpha ({\bf q}, \omega) = \pi_{\alpha \tilde \alpha}({\bf q}, \omega) E_\alpha ({\bf q},\omega)
\nonumber \\
&& \hspace{3mm}
= \frac{1}{V} \sum_{LL'}\sum_{{\bf k}\sigma} 
J_\alpha^{LL'}({\bf k},{\bf k}_+) \langle c^\dagger_{L{\bf k} \sigma} c_{L'{\bf k}_+{\bf q} \sigma} \rangle_\omega
\nonumber \\
&& \hspace{3mm}
= \frac{1}{V} \sum_{LL'}\sum_{{\bf k}\sigma} 
J_\alpha^{LL'}({\bf k},{\bf k}_+) \delta n^{LL'} ({\bf k}, {\bf q}, \omega).
\label{eq32}
\end{eqnarray}
	Similarly, the induced charge density is given by 
\begin{eqnarray}
&& \hspace{-10mm}
J_0 ({\bf q}, \omega) =  \frac{1}{V} \sum_{LL'}\sum_{{\bf k}\sigma} 
J_0^{LL'}({\bf k},{\bf k}_+) \delta n^{LL'} ({\bf k}, {\bf q}, \omega).
\label{eq33}
\end{eqnarray}

\section{Bethe-Salpeter expressions for $\pi_{\mu \nu}^{\rm intra} ({\bf q},{\it i} \nu_n)$}
Let us now restrict our attention to
a single-band case and explain how the simultaneous treatment of the Dyson equation, the Bethe-Salpeter equations, 
and the charge continuity equation mentioned in Sec.~IV\,A works in typical approximate schemes.
	In this paper, the correlation functions $\pi_{\mu \nu} ({\bf q},{\it i} \nu_n)$ are shown in terms of ${\cal G} ({\bf k}, {\it i}\omega_n)$ and 
$\Phi_{\nu}({\bf k},{\bf k}_+,{\it i}\omega_n,{\it i}\omega_{n+})$, and instead of the Bethe-Salpeter equation 
for $\Phi_{\nu}({\bf k},{\bf k}_+,{\it i}\omega_n,{\it i}\omega_{n+})$, 
we use the corresponding quantum kinetic equation
\begin{eqnarray}
&& \hspace{-5mm}
\big[{\it i} \hbar \nu_n + \varepsilon_0({\bf k},{\bf k}_+) \big] 
\Phi_{\nu} ({\bf k},{\bf k}_+, {\it i} \omega_n, {\it i} \omega_{n+} )
\nonumber \\
&&  \hspace{0mm}
= \frac{1}{\hbar} \big[{\cal G} ({\bf k}, {\it i} \omega_n)  - {\cal G} ({\bf k}_+, {\it i} \omega_{n+})\big] 
J_\nu ({\bf k}_+,{\bf k})
\nonumber \\
&&  \hspace{-5mm} 
-\lambda^2 \big[\hbar \Sigma({\bf k}, {\it i} \omega_n )- \hbar \Sigma({\bf k}_+, {\it i} \omega_{n+} ) \big]
\Phi_{\nu} ({\bf k},{\bf k}_+, {\it i} \omega_n, {\it i} \omega_{n+})
\nonumber \\
&&  \hspace{0mm} 
-\lambda^2\frac{1}{\hbar}\big[ {\cal G}({\bf k}, {\it i}\omega_n)- {\cal G}({\bf k}_+, {\it i}\omega_{n+}) \big]
\nonumber \\
&&  \hspace{5mm} 
\times \sum_{{\bf k}' \sigma'}\frac{1}{\beta} \sum_{{\it i} \omega_m}
\Phi_{\nu} ({\bf k}',{\bf k}'_+, {\it i} \omega_m, {\it i} \omega_{m+})
\nonumber \\
&&  \hspace{10mm} 
\times  U ({\bf k}_+,{\bf k}',{\bf k}'_+, {\bf k}, {\it i} \omega_{n+}, {\it i} \omega_{m}
{\it i} \omega_{m+}, {\it i} \omega_n)
\label{eq34}
\end{eqnarray}
[$\varepsilon_0({\bf k},{\bf k}_+) = \varepsilon_0({\bf k}) - \varepsilon_0({\bf k}_+)$].
	This equation is equivalent to the original Bethe-Salpeter equation, and also represents the generalization 
of the intraband part of self-consistent equation (\ref{eq30}).
	This equation is an integral equation of a complicated kind.
	For simplicity we omit here explicit reference to the conduction band index.

According to the first expression in the third row of Fig.~1, the correlation functions $\pi_{\mu \nu} ({\bf q},{\it i} \nu_n)$ 
can be shown in the following way \cite{Schrieffer64,Ziman88,Kupcic15}
\begin{eqnarray}
&& \hspace{-5mm}
\pi_{\mu \nu} ({\bf q}, {\it i}  \nu_n) 
= \frac{1}{V} \sum_{{\bf k} \sigma} J_\mu ({\bf k},{\bf k}_+)  \frac{1}{\beta \hbar^2} \sum_{{\it i}\omega_n} {\cal G} ({\bf k}, {\it i}\omega_n) 
\nonumber \\
&&  \hspace{15mm}
\times  {\cal G} ({\bf k}_+, {\it i}\omega_{n+}) \Gamma_\nu({\bf k}_+,{\bf k},{\it i}\omega_{n+}, {\it i}\omega_{n}) 
\label{eq35} \\
&& \hspace{13mm}
= \frac{1}{V} \sum_{{\bf k} \sigma} J_\mu({\bf k},{\bf k}_+) 
\frac{1}{\beta} \sum_{{\it i}\omega_n} \Phi_{\nu} ({\bf k},{\bf k}_+,{\it i}\omega_n,{\it i}\omega_{n+}).
\nonumber \\
\label{eq36}
\end{eqnarray}
	The relation between the renormalized vertex function $\Gamma_\nu({\bf k}_+,{\bf k},{\it i}\omega_{n+}, {\it i}\omega_{n}) $
and the electron-hole propagator $\Phi_{\nu} ({\bf k},{\bf k}_+,{\it i}\omega_n,{\it i}\omega_{n+})$ is thus
\begin{eqnarray}
&& \hspace{-10mm}
\Phi_{\nu} ({\bf k},{\bf k}_+,{\it i}\omega_n,{\it i}\omega_{n+}) =
\frac{1}{\hbar^2} {\cal G} ({\bf k}, {\it i}\omega_n) {\cal G} ({\bf k}_+, {\it i}\omega_{n+}) 
\nonumber \\
&&  \hspace{25mm}
\times  \Gamma_\nu({\bf k}_+,{\bf k},{\it i}\omega_{n+}, {\it i}\omega_{n}).
\label{eq37} 
\end{eqnarray}
	On the other hand, the second expression in the third row leads to 
\begin{eqnarray}
&& \hspace{-5mm}
\pi_{\mu \nu} ({\bf q}, {\it i}  \nu_n) 
= \frac{1}{V} \sum_{{\bf k} \sigma} \frac{1}{\beta} \sum_{{\it i}\omega_n} \Phi_{\mu} ({\bf k}_+,{\bf k},{\it i}\omega_{n+},{\it i}\omega_n)
J_\nu({\bf k}_+,{\bf k}).
\nonumber \\
\label{eq38}
\end{eqnarray}

For long wavelengths,
the charge vertex $J_0({\bf k}_+,{\bf k}) \approx e$ is a constant and the current vertex $J_\alpha({\bf k}_+,{\bf k}) \approx e v_\alpha({\bf k})$
is proportional to the electron group velocity $v_\alpha({\bf k})$.
	This means that the electron-hole propagator $\Phi_{\nu} ({\bf k},{\bf k}_+,{\it i}\omega_n,{\it i}\omega_{n+})$ can be shown 
as a sum of four contributions of different symmetries,
\begin{eqnarray}
&& \hspace{-5mm}
\Phi_{\nu} ({\bf k},{\bf k}_+,{\it i}\omega_n,{\it i}\omega_{n+}) = \sum_{\mu'=0,x,y,z} \Phi_{\nu[\mu']} ({\bf k},{\bf k}_+,{\it i}\omega_n,{\it i}\omega_{n+}) ,
\nonumber \\
\label{eq39}
\end{eqnarray}
where 
$\Phi_{\nu[\mu']} ({\bf k},{\bf k}_+,{\it i}\omega_n,{\it i}\omega_{n+}) \propto J_{\mu'}({\bf k}_+,{\bf k}) \approx  J_{\mu'}({\bf k})$,
resulting in
\begin{eqnarray}
&& \hspace{-10mm}
\pi_{\mu \nu} ({\bf q}, {\it i}  \nu_n) 
\nonumber \\
&&  \hspace{-5mm}
=
\frac{1}{V} \sum_{{\bf k} \sigma} J_\mu({\bf k},{\bf k}_+) 
\frac{1}{\beta} \sum_{{\it i}\omega_n} \Phi_{\nu[\mu]} ({\bf k},{\bf k}_+,{\it i}\omega_n,{\it i}\omega_{n+})
\nonumber \\
&&  \hspace{-5mm}
= \frac{1}{V} \sum_{{\bf k} \sigma} \frac{1}{\beta} \sum_{{\it i}\omega_n} \Phi_{\mu[\nu]} ({\bf k}_+,{\bf k},{\it i}\omega_{n+},{\it i}\omega_n)
J_\nu({\bf k}_+,{\bf k}).
\label{eq40}
\end{eqnarray}
	Evidently for electromagnetic fields polarized along the $\alpha$ axis, there are only two components in Eq.~(\ref{eq39}), i.e.,
\begin{eqnarray}
&& \hspace{-5mm}
\Phi_{\nu} ({\bf k},{\bf k}_+,{\it i}\omega_n,{\it i}\omega_{n+}) = \sum_{\mu'=0, \alpha} \Phi_{\nu[\mu']} ({\bf k},{\bf k}_+,{\it i}\omega_n,{\it i}\omega_{n+}).
\nonumber \\
\label{eq41}
\end{eqnarray}

First important consequence of Eq.~(\ref{eq40}) is that the charge continuity equation from Eq.~(\ref{eq14}) can be shown in the following way
\begin{eqnarray}
&&  \hspace{-10mm}
\frac{1}{V} \sum_{{\bf k} \sigma} \sum_{\mu=0,x,y,z} q_\mu J_\mu({\bf k},{\bf k}_+) \Phi_{0[\mu]} ({\bf k},{\bf k}_+,\omega) = 0.
\label{eq42}
\end{eqnarray}
	Here, $\Phi_{\nu} ({\bf k},{\bf k}_+,\omega)$ is the analytically continued form of
\begin{eqnarray}
&& \hspace{-5mm}
\Phi_{\nu} ({\bf k},{\bf k}_+,{\it i} \nu_n) = \frac{1}{\beta} \sum_{ {\it i} \omega_n} 
\Phi_{\nu} ({\bf k},{\bf k}_+, {\it i} \omega_n, {\it i} \omega_{n+}),
\label{eq43}
\end{eqnarray}
and the $q_\mu$ are the components of the four-component wave vector $q = (\omega, {\bf q})$.
	Similarly, the charge continuity equation from Eq.~(\ref{eq15}) leads to
\begin{eqnarray}
&&  \hspace{-10mm}
\frac{1}{V} \sum_{{\bf k} \sigma} \sum_{\mu=0,x,y,z} q_\mu J_\mu({\bf k},{\bf k}_+) \Phi_{\alpha[\mu]} ({\bf k},{\bf k}_+,\omega) 
\nonumber \\
&&  \hspace{0mm}
=  \frac{1}{V} \sum_{{\bf k} \sigma} \sum_{\beta=x,y,z} q_\beta J_\beta({\bf k},{\bf k}_+) \Phi_{\alpha[\beta]} ({\bf k},{\bf k}_+,0).
\label{eq44}
\end{eqnarray}
	Finally, it is important to notice that the same symmetry based analysis holds for the intraband contributions in Sec.~IV\,B as well.
	The relation between the two notations is the following 
\begin{eqnarray}
&& \hspace{-5mm}
\delta n ({\bf k}, {\bf q}, \omega) =
({\it i}/q_\alpha) \Phi_{0} ({\bf k},{\bf k}_+,\omega) E_\alpha ({\bf q},\omega).
\label{eq45}
\end{eqnarray}

\subsection{Common Fermi liquid theory}
In the Landau theory of Fermi liquids \cite{Pines89,Platzman73}, 
electrodynamic properties of conduction electrons are described by the conductivity tensor
\begin{eqnarray}
&& \hspace{-10mm}
\sigma_{\alpha \alpha} ({\bf q}, \omega) = \frac{{\it i}}{q_\alpha}  \pi_{\alpha 0} ({\bf q}, \omega)
\equiv \pi_{\alpha \tilde \alpha} ({\bf q}, \omega)
\nonumber \\
&&  \hspace{5mm}
= \frac{1}{V} \sum_{{\bf k} \sigma} J_\alpha({\bf k},{\bf k}_+) \frac{{\it i}}{q_\alpha} \Phi_{0[\alpha]} ({\bf k},{\bf k}_+,\omega).
\label{eq46}
\end{eqnarray}
	Here, $\Phi_{0[\alpha]} ({\bf k},{\bf k}_+,\omega)$ is the solution of the Landau-Silin kinetic equation,
which is simplified version of the equations (\ref{eq30}) and (\ref{eq34}) \cite{Platzman73,Kupcic15}.
	In this theory, the main simplification is in the way how vertex corrections are taken into account.
	Namely, for electromagnetic fields polarized along the $\alpha$ axis, we can insert the assumption (\ref{eq41})
into Eq.~(\ref{eq34}), separate all contributions which are odd functions of $k_\alpha$ from the even contributions, 
and use the ansatz for the sum of the second and third term on the right-hand side of the kinetic equation which makes the sum of 
the even contributions identical to the charge continuity equation (\ref{eq42}).
	In this way, Eq.~(\ref{eq34}) reduces to two coupled equations for $\Phi_{0[0]} ({\bf k},{\bf k}_+,\omega)$ and
$\Phi_{0[\alpha]} ({\bf k},{\bf k}_+,\omega)$; the first one is the charge continuity equation and the second one is the 
transport equation \cite{Pines89,Kupcic14}.
	After retaining only the leading contributions to the self-energy $\Sigma({\bf k},{\it i}\omega_n)$ and the related 
contributions to the irreducible four-point interaction, we obtain the well-known textbook expression 
for $\sigma_{\alpha \alpha} ({\bf q}, \omega)$.
	This conductivity formula is known to describe well the Thomas-Fermi static screening, the collective modes of the electronic 
subsystem as well as the dc and dynamical conductivity.

Let us now present the formal derivation of
both the memory-function conductivity formula [Eq.~(\ref{eq51})]
and its simplified form in which the issue of the Thomas-Fermi static screening is taken aside [Eq.~(\ref{eq54})].
	These expressions reduce to the well-known Fermi-liquid expressions when the memory function $M_\alpha ({\bf k},\omega)$
is approximated by its imaginary part $M_\alpha^i ({\bf k},\omega) \approx \Gamma_\alpha ({\bf k})$
[here $\Gamma_\alpha ({\bf k})$ is the usual notation for the relaxation rate, which depends on ${\bf k}$ 
and on the polarization index $\alpha$].

\section{Memory-function conductivity formula}
The present derivation of the memory-function conductivity formula follows the same general path as
the textbook derivation of the transport coefficients in the Fermi liquid theory 
\cite{Pines89}.
	We consider the quantum kinetic equation for $\Phi_0 ({\bf k},{\bf k}_+,{\it i}\omega_n,{\it i}\omega_{n+})$
in the presence of the electromagnetic field polarized along the $\alpha$ axis, and use the ansatz
\begin{eqnarray}
&& \hspace{-5mm}
- \lambda^2 \hbar  \Pi ({\bf k},{\bf k}_+,  {\it i}\omega_{n}, {\it i}\omega_{n+})
\Phi_{0[\alpha]} ({\bf k},{\bf k}_+, {\it i} \omega_n, {\it i} \omega_{n+} )
\label{eq47}
\end{eqnarray}
for the sum of the last two terms on the right-hand side of the equation.
	In this way, this integral equation transforms into an ordinary equation
\begin{eqnarray}
&& \hspace{-10mm}
\big[{\it i} \hbar \nu_n + \varepsilon_0({\bf k},{\bf k}_+) \big] 
\Phi_0 ({\bf k},{\bf k}_+, {\it i} \omega_n, {\it i} \omega_{n+} )
\nonumber \\
&&  \hspace{3mm}
+ \hbar  \Pi ({\bf k},{\bf k}_+,  {\it i}\omega_{n}, {\it i}\omega_{n+}) \big]
\Phi_{0[\alpha]} ({\bf k},{\bf k}_+, {\it i} \omega_n, {\it i} \omega_{n+} )
\nonumber \\
&&  \hspace{0mm}
= \frac{1}{\hbar}  \big[{\cal G} ({\bf k}, {\it i} \omega_n)  - {\cal G} ({\bf k}_+, {\it i} \omega_{n+})\big] 
J_{0} ({\bf k}_+,{\bf k}).
\label{eq48}
\end{eqnarray}
	It is easily seen that summation over ${\bf k}$ and ${\it i} \omega_n$ leads to Eq.~(\ref{eq42}).
	Therefore, the ansatz (\ref{eq47}) is consistent with the charge continuity equation.
	Here, $\Pi({\bf k},{\bf k}_+, {\it i}\omega_{n},{\it i}\omega_{n+})
=  \widetilde \Sigma ({\bf k}, {\it i}\omega_n) - \widetilde \Sigma ({\bf k}_+, {\it i}\omega_{n+})$ is the electron-hole self-energy
and the unknown quantity $\widetilde \Sigma ({\bf k}, {\it i}\omega_n)$ is the modified single-electron self-energy.

The next level of approximation is to
replace the electron-hole self-energy $\Pi ({\bf k},{\bf k}_+,  {\it i}\omega_{n}, {\it i}\omega_{n+})$ 
by the quantity which depends only on the difference of two electron frequencies and 
on the direction of the wave vector ${\bf q} = q_\alpha \hat e_\alpha$
[that is, $\Pi ({\bf k},{\bf k}_+,  {\it i}\omega_{n}, {\it i}\omega_{n+}) \approx
M_\alpha({\bf k}, {\it i} \nu_n )$].
	Then the kinetic equation becomes
\begin{eqnarray}
&& \hspace{-10mm}
\big[{\it i} \hbar \nu_n + \varepsilon_0({\bf k},{\bf k}_+) \big] \Phi_{0} ({\bf k},{\bf k}_+, {\it i} \omega_n, {\it i} \omega_{n+} )
\nonumber \\
&&  \hspace{3mm}
+ \hbar M_\alpha({\bf k}, {\it i} \nu_n )
\Phi_{0[\alpha]} ({\bf k},{\bf k}_+, {\it i} \omega_n, {\it i} \omega_{n+} )
\nonumber \\
&&  \hspace{0mm}
= \frac{1}{\hbar}   \big[{\cal G} ({\bf k}, {\it i} \omega_n)  - {\cal G} ({\bf k}_+, {\it i} \omega_{n+})\big] 
J_{0} ({\bf k}_+,{\bf k}).
\label{eq49}
\end{eqnarray}
	Summation over  ${\it i} \omega_n$ is straightforward now.
	After using the momentum distribution function $n({\bf k})$ from Eq.~(\ref{eq25}) and the electron-hole propagator
$\Phi_{0} ({\bf k},{\bf k}_+,\omega) = \sum_{\mu'=0, \alpha} \Phi_{0[\mu']} ({\bf k},{\bf k}_+,\omega)$ from Eq.~(\ref{eq43}),
we obtain
\begin{eqnarray}
&& \hspace{-5mm}
q_\alpha v_\alpha ({\bf k}) \Phi_{0[0]} ({\bf k},{\bf k}_+, \omega)
- [\omega + M_\alpha({\bf k}, \omega ) ] \Phi_{0[\alpha]} ({\bf k},{\bf k}_+, \omega)
\nonumber \\
&&  \hspace{3mm}
+  (e/\hbar) [n({\bf k})  - n({\bf k}_+)]  
\nonumber \\
&&  \hspace{0mm}
= \omega \Phi_{0[0]} ({\bf k},{\bf k}_+, \omega) 
- q_\alpha v_\alpha ({\bf k}) \Phi_{0[\alpha]} ({\bf k},{\bf k}_+, \omega).
\label{eq50}
\end{eqnarray}
	This equation is decomposed into the odd contributions and the even contributions.
	The right-hand side and the left-hand side expressions must vanish independently, and we obtain two equations, 
which can be easily solved, for example, for $\Phi_{0[\alpha]} ({\bf k},{\bf k}_+,\omega)$.
	By substituting this expression for $\Phi_{0[\alpha]} ({\bf k},{\bf k}_+,\omega)$ into the definition relation (\ref{eq46}), 
we obtain the memory-function conductivity formula
\begin{eqnarray}
&& \hspace{-5mm} 
\sigma_{\alpha \alpha} ({\bf q}, \omega) =
\frac{1}{V} \sum_{{\bf k} \sigma}
{\it i} \hbar |J_{\alpha}({\bf k},{\bf k}_+)|^2
\frac{n({\bf k})-n({\bf k}_+)}{\varepsilon_0({\bf k}_+,{\bf k})}
\nonumber \\
&& \hspace{5mm}
\times \frac{\hbar \omega}{\hbar \omega(\hbar \omega + 
\hbar M_\alpha ({\bf k},\omega)) - \varepsilon^2_0({\bf k},{\bf k}_+)}.
\label{eq51}
\end{eqnarray}
	The function $M_\alpha ({\bf k},\omega)$ is usually called the memory function.

In the static limit, the result is the static Thomas-Fermi dielectric susceptibility
\begin{eqnarray}
&& \hspace{-10mm} 
- 4 \pi \pi_{00} ({\bf q}) = 4 \pi e^2 
\frac{1}{V} \sum_{{\bf k} \sigma} \frac{n({\bf k})-n({\bf k}_+)}{\varepsilon_0({\bf k}_+,{\bf k})} \equiv k_{\rm TF}^2,
\label{eq52}
\end{eqnarray}
which is proportional to the density of states at the Fermi level
\begin{eqnarray}
&& \hspace{-10mm} 
\rho (\mu) = \frac{1}{V} \sum_{{\bf k} \sigma} \bigg(-\frac{\partial n({\bf k})}{\partial \varepsilon_0({\bf k})}\bigg),
\label{eq53}
\end{eqnarray}
as well as to the square of the Thomas-Fermi wave vector $k_{\rm TF}$.
	In the Drude limit $(\hbar \omega)^2 \gg \varepsilon^2_0({\bf k},{\bf k}_+)$, on the other hand, the result is
\begin{eqnarray}
&& \hspace{-10mm}
\sigma_{\alpha \alpha} ({\bf q},\omega) = 
\frac{{\it i} e^2}{m}\frac{1}{V} \sum_{{\bf k} \sigma} \bigg(-\frac{\partial n({\bf k})}{\partial \varepsilon_0({\bf k})}\bigg)
\frac{m v_\alpha^2 ({\bf k})}{\omega  +  \lambda^2 M_\alpha({\bf k},\omega)}
\nonumber \\
&&  \hspace{5mm}
= \frac{{\it i} e^2}{m}\frac{1}{V} \sum_{{\bf k} \sigma} m v_\alpha^2 ({\bf k}) \bigg(-\frac{\partial n({\bf k})}{\partial \varepsilon_0({\bf k})}\bigg)
\frac{1}{\omega + {\it i} \eta}  
\nonumber \\
&&  \hspace{8mm}
 \times \bigg(1 - \lambda^2 \frac{M_\alpha({\bf k},\omega)}{\omega} + \cdots \bigg).
\label{eq54}
\end{eqnarray}
	It must be emphasized that
in the Drude limit the same result follows after using the approximation 
$\Phi_{0} ({\bf k},{\bf k}_+, {\it i} \omega_n, {\it i} \omega_{n+} ) \approx
\Phi_{0[\alpha]} ({\bf k},{\bf k}_+, {\it i} \omega_n, {\it i} \omega_{n+} )$ in Eq.~(\ref{eq49}).
	This type of approximation is widely used in the textbook discussions of the transport equations
\cite{Pines89,Platzman73,Ziman72}.

It is also important to notice that the $\lambda^0$ term in Eq.~(\ref{eq54})
describes the ideal conductivity, i.e., ${\rm Re} \{\sigma_{\alpha \alpha}^{(0)} ({\bf q},\omega) \} \propto \delta (\omega)$, 
and that the corresponding integrated spectral weight is in agreement 
with the intraband part of the sum rule (\ref{eq20}).
	The corrections, starting with the $\lambda^2$ term, lead to the redistribution of the spectral weight 
over a wide frequency range (see Sec.~VIII\,A).

\begin{figure}
   \centerline{\includegraphics[width=20pc]{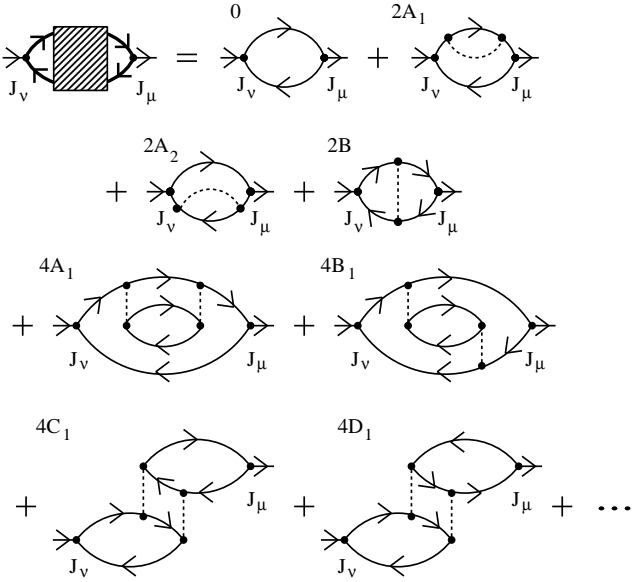}}
   \caption{The expansion of the correlation function $\pi_{\mu \nu} ({\bf q}, {\it i} \nu_n)$ in powers of $\lambda'$ 
   [$\lambda'$ is the perturbation parameter in $H' = \lambda' H'_1 + (\lambda')^2 H_2'$] \cite{Kupcic14,Kupcic15}.
   }
  \end{figure}

\subsection{Low-order perturbation theory}
The usual way to determine the structure of the memory function $M_\alpha({\bf k},\omega)$ 
is to compare the expansion of the conductivity tensor (\ref{eq54}) in powers of $\lambda^2$
with the usual (low-order) perturbation expansion of $\pi_{\alpha 0} ({\bf k}, \tau)$ from Eq.~(\ref{eq26}) 
in powers of $\lambda'$ \cite{Kupcic15,Kupcic16}.
	Here, $\lambda'$ is the perturbation parameter in the perturbation $H' = \lambda' H'_1 + (\lambda')^2 H_2'$, and 
$H_1' = H'_{1a} + H'_{1b}$.

The evaluation of $M_{\alpha} ({\bf k}, \omega)$ is very difficult in general.
	However, to obtain the leading terms from the common Fermi liquid theory, it suffices to work out
the diagrams shown in Fig.~2 and identify the function $M_{\alpha} ({\bf k}, \omega)$, which is now a complex function of $\omega$.
	The results are \cite{Kupcic14,Kupcic15}
\begin{eqnarray}
&& \hspace{-10mm}
\hbar M^{[2]}_{\alpha} ({\bf k}, \omega) = -\frac{1}{N} \sum_{\lambda {\bf k'}} 
|G_\lambda({\bf k},{\bf k}')|^2 \bigg(1 -\frac{v_\alpha ({\bf k}')}{v_\alpha ({\bf k})} \bigg) 
\nonumber \\
&& \hspace{-5mm}
\times \sum_{s = \pm 1} \sum_{s' = \pm 1} \frac{s'\big[f^b (s'\omega_{\lambda {\bf k}-{\bf k}'}) 
+ f(s \varepsilon_0({\bf k}')) \big]}{
\hbar \omega + {\it i} \eta + s \varepsilon_0({\bf k},{\bf k}') + s' \hbar \omega_{\lambda {\bf k}-{\bf k}'}}
\nonumber \\
&& \hspace{8mm}
= -\frac{1}{N} \sum_{\lambda {\bf k'}} 
|G_\lambda({\bf k},{\bf k}')|^2 \bigg(1 -\frac{v_\alpha ({\bf k}')}{v_\alpha ({\bf k})} \bigg) 
\nonumber \\
&& \hspace{-5mm}
\times \sum_{s = \pm 1} \sum_{s' = \pm 1} \frac{f^b (\omega_{\lambda {\bf k}-{\bf k}'}) 
+ f(s s' \varepsilon_0({\bf k}'))}{
\hbar \omega + {\it i} \eta + s \varepsilon_0({\bf k},{\bf k}') + s' \hbar \omega_{\lambda {\bf k}-{\bf k}'}}
\nonumber \\
&& \hspace{8mm}
= -\frac{1}{N} \sum_{\lambda {\bf k'}} 
|\widetilde G_\lambda({\bf k},{\bf k}')|^2 {\cal S}({\bf k}, {\bf k}',\omega) 
\label{eq55}
\end{eqnarray}
and 
\begin{eqnarray}
&& \hspace{-5mm}
\hbar M^{[4]}_{\alpha} ({\bf k},\omega) = 
-\sum_{{\bf k'}{\bf q} \sigma'} \frac{|\varphi_{\sigma \sigma'}({\bf q})|^2}{V^2} 
 \frac{1}{v_\alpha ({\bf k})}\big[v_\alpha ({\bf k}) + v_\alpha ({\bf k}'_+)
\nonumber \\
&& \hspace{5mm}
- v_\alpha ({\bf k}') -v_\alpha ({\bf k}_+)  \big][f(\varepsilon_0({\bf k}'))-f(\varepsilon_0({\bf k}'_+))]
\nonumber \\
&& \hspace{0mm}
\times \sum_{s = \pm 1} \frac{f^{b}(\omega({\bf k}'_+,{\bf k}')) + f(\varepsilon_0({\bf k}_+))}{
\hbar \omega + {\it i} \eta + s \varepsilon_{0}({\bf k},{\bf k}') + s \varepsilon_{0}({\bf k}'_+,{\bf k}_+)}
\label{eq56}
\end{eqnarray}
for the scattering by phonons and by other electrons
[the indices $[2]$ and $[4]$ stand for $(\lambda')^2$ and $(\lambda')^4$, respectively].
	Here, $f^{b}(\omega)$ is the Bose-Einstein distribution function, and 
$\hbar \omega({\bf k},{\bf k}') = \varepsilon_{0}({\bf k})-\varepsilon_{0}({\bf k}')$.
	It should be noted that the $(H'_{1b})^2$ contribution originating from the scattering by static disorder is described by 
Eq.~(\ref{eq55}) as well.
	The result for these scattering processes (labeled by the index $\lambda = 0$) is found by taking the limit
$\omega_{0{\bf k}'-{\bf k}} \rightarrow 0$ and 
$|G_0({\bf k},{\bf k}')|^2\sum_{s'} s'f^b (s'\omega_{0 {\bf k}'-{\bf k}}) \rightarrow | V({\bf k},{\bf k}')|^2$.

As shown in Ref.~\cite{Kupcic09} and \cite{Kupcic15}, this form of the memory function $M_\alpha({\bf k},{\it i} \nu_n)$
can be obtained from $\Pi ({\bf k},{\bf k}_+,  {\it i}\omega_{n}, {\it i}\omega_{n+})$
by measuring the energy of the electron in $\widetilde \Sigma ({\bf k}_+, {\it i}\omega_{n+})$ [hole in $\widetilde \Sigma ({\bf k}, {\it i}\omega_n)$]
with respect to the energy of hole (electron), and not with respect to the chemical potential $\mu$; that is
\begin{eqnarray} 
&& \hspace{-10mm}
M_\alpha ({\bf k}, {\it i} \nu_n) =
\widetilde \Sigma ({\bf k},\varepsilon_0({\bf k}_+)/\hbar- {\it i} \nu_n)
\nonumber \\
&&\hspace{13mm}
- \widetilde \Sigma ({\bf k}_+,\varepsilon_0({\bf k})/\hbar+ {\it i} \nu_n).
\label{eq57}
\end{eqnarray}
	This relation, together with Eqs.~(\ref{eq55}) and (\ref{eq56}), illustrates how 
the modified self-energy $\widetilde \Sigma ({\bf k}, {\it i}\omega_n)$ is related to the single-electron
self-energy $\Sigma ({\bf k}, {\it i}\omega_n)$.
	For example, for the scattering by phonons, the $(\lambda')^2$ term follows after replacing the coupling constant 
$|G_\lambda({\bf k},{\bf k}')|^2$ in 
\begin{eqnarray}
&& \hspace{-10mm}
\hbar \Sigma^{[2]} ({\bf k}, \omega)  = \frac{1}{N} \sum_{\lambda {\bf k'}}  |G_\lambda({\bf k},{\bf k}')|^2 
\nonumber \\
&& \hspace{5mm}
\times \sum_{s = \pm 1} \frac{f^b (\omega_{\lambda {\bf k}'-{\bf k}}) + f(s\varepsilon_0({\bf k}'))}{
\hbar \omega  + {\it i} \eta  - \varepsilon_0({\bf k}') + \mu + s \hbar \omega_{\lambda {\bf k}'-{\bf k}}}
\label{eq58}
\end{eqnarray}
by 
$|\widetilde G_\lambda({\bf k},{\bf k}')|^2 = |G_\lambda({\bf k},{\bf k}')|^2 (1 -v_\alpha ({\bf k}')/v_\alpha ({\bf k}))$
and the chemical potential $\mu$ in the denominator by $\varepsilon_0({\bf k})$.

The extra factor $(1 -v_\alpha ({\bf k}')/v_\alpha ({\bf k}))$ in Eq.~(\ref{eq55}) causes a reduction of
the forward scattering contributions (${\bf k}'\approx {\bf k}$) in $\widetilde \Sigma ({\bf k}, {\it i}\omega_n)$
with respect to $\Sigma ({\bf k}, {\it i}\omega_n)$.
	A direct consequence of
this effect is the fact that the intraband memory-function conductivity formula is characterized by two different damping energies;
the first one, $\Sigma^i ({\bf k}, \omega)$ in $n({\bf k})$, describes the lifetime of the electron, and the second one,
$\widetilde \Sigma^i ({\bf k}, \omega)$ in ${\rm Im} \{ M_\alpha({\bf k},\omega) \}$, the corresponding relaxation time,
with $\Sigma^i ({\bf k}, \omega) > \widetilde \Sigma^i ({\bf k}, \omega)$.

The extra factor $(v_\alpha ({\bf k}) + v_\alpha ({\bf k}'_+)
- v_\alpha ({\bf k}') -v_\alpha ({\bf k}_+))/v_\alpha ({\bf k})$ in $M^{[4]}_{\alpha} ({\bf k},\omega)$ has even stronger effect.
	Evidently in the electron-electron scattering channel, not only the forward scattering contributions but also
the normal backward scattering contributions drop out of the function $M_{\alpha} ({\bf k},\omega)$.
	Only the umklapp backward scattering contributions remain.

\subsection{Self-consistent RPA approach}
To understand the significance of the memory function $M_{\alpha} ({\bf k}, \omega)$ in the language of
the generalized self-consistent RPA equations, let us consider
the same case as in Sec.~IV\,B.
	We must solve the integral equation 
\begin{eqnarray}
&& \hspace{-10mm}
[\hbar \omega + \varepsilon_0({\bf k},{\bf k}_+)  + {\it i} \eta ] \delta n ({\bf k},{\bf q},\omega)
\nonumber \\
&& \hspace{0mm}
= [n({\bf k}_+) - n({\bf k})] P_\alpha ({\bf k}_+,{\bf k})E_\alpha ({\bf q},\omega)
\nonumber \\
&& \hspace{3mm}
+ \sum_{\lambda {\bf k}'}\frac{|G_\lambda({\bf k},{\bf k}')|^2}{N}\big[ {\cal S}({\bf k}, {\bf k}',\omega) \delta n ({\bf k},{\bf q},\omega)
\nonumber \\
&& \hspace{3mm}
- {\cal S}({\bf k}', {\bf k},\omega) \delta n ({\bf k}',{\bf q},\omega) \big],
\label{eq59}
\end{eqnarray}
and insert the resulting expression for $\delta n ({\bf k},{\bf q},\omega)$ into Eq.~(\ref{eq32}) or Eq.~(\ref{eq33}).

First, it is important to realize that
multiplication of Eq.~(\ref{eq59}) by $J_0({\bf k},{\bf k}_+) \approx e$ and summation over ${\bf k}$ leads
to the charge continuity equation from Eq.~(\ref{eq14}).
	This means that the charge continuity equation is satisfied in this case at least on average.

Multiplication by $J_\alpha({\bf k},{\bf k}_+) \approx e v_\alpha ({\bf k})$ and summation over ${\bf k}$ leads to
\begin{eqnarray}
&& \hspace{-10mm}
\sum_{{\bf k}\sigma} e v_\alpha ({\bf k}) \delta n({\bf k},{\bf q},\omega)
= \sum_{{\bf k}\sigma} e v_\alpha ({\bf k}) \bigg( \delta n^{(0)}({\bf k},{\bf q},\omega)
\nonumber \\
&& \hspace{10mm}
-\delta n ({\bf k}, {\bf q}, \omega)
\lambda^2 \frac{\hbar M_{\alpha} ({\bf k}, \omega)}{\hbar \omega + \varepsilon_0({\bf k},{\bf k}_+)  + {\it i} \eta} \bigg),
\label{eq60}
\end{eqnarray}
where
\begin{eqnarray}
&& \hspace{-5mm}
\delta n^{(0)}({\bf k},{\bf q},\omega) = \frac{ n({\bf k}_+) - n({\bf k})}{\hbar \omega + \varepsilon_0({\bf k},{\bf k}_+)  + {\it i} \eta}
P_\alpha ({\bf k}_+,{\bf k})E_\alpha ({\bf q},\omega),
\nonumber \\
\label{eq61}
\end{eqnarray}
$P_\alpha ({\bf k}_+,{\bf k}) =-{\it i} e/q_\alpha$, and
\begin{eqnarray}
&& \hspace{-3mm}
\hbar M_{\alpha} ({\bf k}, \omega) \approx -\frac{1}{N} \sum_{\lambda {\bf k'}} 
|\widetilde G_\lambda({\bf k},{\bf k}')|^2  {\cal S}({\bf k}, {\bf k}',\omega).
\label{eq62}
\end{eqnarray}
	Equation (\ref{eq60}) can most easily be solved
if we show the nonequilibrium distribution function $\delta n ({\bf k}, {\bf q}, \omega)$ in the form
\begin{eqnarray}
&& \hspace{-10mm}
 \delta n ({\bf k}, {\bf q}, \omega) = \sum_{n = 0}^\infty \lambda^{2n} \delta n^{(2n)} ({\bf k}, {\bf q}, \omega),
\label{eq63}
\end{eqnarray}
and recognize a simple recursion relation for the coefficients $\delta n^{(2n)} ({\bf k}, {\bf q}, \omega)$,
\begin{eqnarray}
&& \hspace{-5mm}
\delta n^{(2n+2)} ({\bf k}, {\bf q}, \omega) = -\frac{\hbar M_{\alpha} ({\bf k}, \omega)}{\hbar \omega + \varepsilon_0({\bf k},{\bf k}_+)  + {\it i} \eta}
\delta n^{(2n)} ({\bf k}, {\bf q}, \omega).
\nonumber \\
\label{eq64}
\end{eqnarray}
	The result for the conductivity tensor is again the memory-function conductivity formula (\ref{eq54}).
	
Not surprisingly, the same result follows from Eq.~(\ref{eq59}) if the sum of the last two terms 
on the right-hand side of the equation is replaced by
\begin{eqnarray}
&& \hspace{-5mm}
\sum_{\lambda {\bf k}'}\frac{|\widetilde G({\bf k},{\bf k}')|^2}{N} {\cal S}({\bf k}, {\bf k}',\omega) 
\delta n_{[\alpha]} ({\bf k}, {\bf q}, \omega).
\label{eq65}
\end{eqnarray}
	This ansatz is equivalent to Eq.~(\ref{eq47}).

\section{Generalized Drude model}
When the memory function $M_{\alpha} ({\bf k},\omega)$ in Eq.~(\ref{eq54}) is replaced by its average over the Fermi surface, 
\begin{equation}
M_{\alpha}(\omega) = \frac{1}{n_{\alpha \alpha}^{\rm intra}}\frac{1}{V} \sum_{{\bf k} \sigma} mv_\alpha^2({\bf k}) 
\bigg(- \frac{\partial n({\bf k})}{\partial \varepsilon_0 ({\bf k})}\bigg) M_{\alpha}({\bf k},\omega),
\label{eq66}
\end{equation}
then we obtain the generalized Drude conductivity formula \cite{Kupcic15}
\begin{eqnarray}
&& \hspace{-10mm}
\sigma_{\alpha \alpha} (\omega) = 
\frac{{\it i} e^2}{m}  \frac{n_{\alpha \alpha}^{\rm intra}}{\omega + {\it i} \eta}  
\bigg(1 - \lambda^2 \frac{M_\alpha(\omega)}{\omega} + \cdots \bigg)
\nonumber \\
&&  \hspace{2mm}
= \frac{{\it i} e^2}{m} \frac{n_{\alpha \alpha}^{\rm intra}}{\omega  +  \lambda^2 M_\alpha(\omega)}.
\label{eq67}
\end{eqnarray}
	Here, 
\begin{eqnarray}
&& \hspace{-10mm}  n_{\alpha \alpha}^{\rm intra} = 
 \frac{1}{V} \sum_{{\bf k} \sigma} mv^2_\alpha({\bf k})\bigg(-\frac{\partial n({\bf k})}{\partial \varepsilon_0({\bf k})}\bigg)
\label{eq68}
\end{eqnarray}
is the intraband contribution to the effective number of charge carriers (\ref{eq22}).
	Evidently in weakly interacting isotropic or nearly isotropic systems
the dependence of $M_{\alpha} ({\bf k},\omega)$ on ${\bf k}$ can be neglected.
	In this case, there is no difference between the two conductivity formulas.
	To obtain $M_\alpha(\omega)$ in this case, it is sufficient to calculate $M_{\alpha} ({\bf k},\omega)$
at an appropriate point at the Fermi surface; 
$M_\alpha(\omega)\approx M_{\alpha} ({\bf k}_{\rm F},\omega)$.

An alternative derivation of the generalized Drude conductivity formula is given in Appendix A.
	The results for the imaginary parts of $M^{[2]}_{\alpha} (\omega)$ and $M^{[4]}_{\alpha} (\omega)$ obtained in this way 
are directly related to the results of the common variational approach for the relaxation rates \cite{Gotze72,Ziman72}.
	These two expressions represent an oversimplified (semiclassical) form of Eqs.~(\ref{eq55}) and (\ref{eq56}).
	Most importantly, the $f(s s' \varepsilon_0({\bf k}'))$ term from the numerator of Eq.~(\ref{eq55}) is missing.
	For example, this means that the scattering by soft phonons is characterized by the factor 
$f^b (\omega_{\lambda {\bf q}'}) + 1/2$ in Eq.~(\ref{eq55}) and by the factor $f^b (\omega_{\lambda {\bf q}'})$ 
in Eq.~(\ref{eqA11}).
	The problem of missing $1/2$ is typical of 
the semiclassical approaches in which the relaxation processes are described in terms of the collision integral.

 \begin{figure}
   \centerline{\includegraphics[width=17pc]{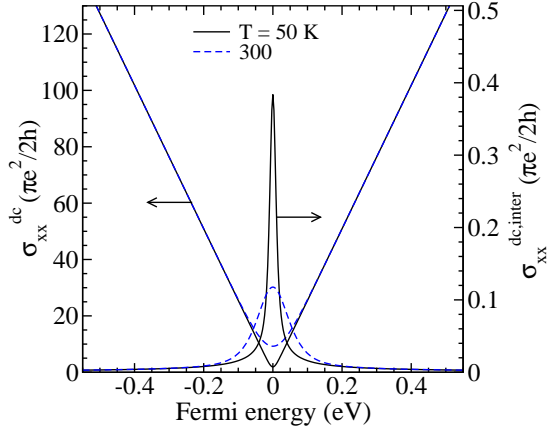}}
   \caption{(Color online) The dependence of the two-band dc conductivity $\sigma_{\alpha \alpha}^{\rm dc}$ from Ref.~\cite{Kupcic16} 
   on the Fermi energy $E_{\rm F}$ ($\equiv \mu$) in the Dirac cone approximation at $T = 50$ and 300 K.
   The solid and dashed lines represent the results of the current-dipole conductivity formula in the relaxation time approximation, for
   $\hbar \Gamma_1 = \hbar \Gamma_2 = 5$ meV and $n_L ({\bf k}) = f_L ({\bf k})$.
   The interband contribution is also shown.
}
  \end{figure}

\section{Intraband memory function in heavily doped graphene}

It is useful first to show the dc limit of the two-band conductivity formula from Ref.~\cite{Kupcic16}
calculated in the relaxation-time approximation. 
	The intraband part is given by Eq.~(\ref{eq54}) with $M_\alpha ({\bf k}, \omega) \approx {\it i} \Gamma_1$.
	The results are shown in Fig.~3 in the doping range $-0.014 < V_0 n < 0.014$ 
(corresponding to $-0.5$ eV $< E_{\rm F} < 0.5$ eV), for both the intraband relaxation rate
$\hbar \Gamma_1$ and the interband relaxation rate $\hbar \Gamma_2$ equal to 5 meV, and for $n_L ({\bf k}) = f_L ({\bf k}) \equiv f(\varepsilon^0_L ({\bf k}))$
($V_0$ is here the primitive cell volume). 
	It is obvious that for $|E_{\rm F}| > 0.1$ the intraband contribution to $\sigma_{\alpha \alpha}(\omega)$ 
is well separated from the interband contribution, and, consequently, $\sigma_{\alpha \alpha}^{\rm dc} \approx \sigma_{\alpha \alpha}^{\rm dc,intra}$.

Therefore, in the heavily doped regime in graphene
(for the Fermi energy $|E_{\rm F}|$ of the order of 0.5 eV or larger), the low-energy conductivity 
$\sigma_{\alpha \alpha} ({\bf q},\omega)$ can be represented by Eq.~(\ref{eq54}) [or by Eq.~(\ref{eq67}), 
in the leading approximation].
	In this case, the dispersion of conduction electrons is 
$\varepsilon_0 ({\bf k}) = \varepsilon_{\pi^*}^0 ({\bf k})$ in the electron doped case and 
$\varepsilon_0 ({\bf k}) = \varepsilon_{\pi}^0 ({\bf k})$ in the hole doped case, where  \cite{Castro09}
\begin{eqnarray}
&& \hspace{-5mm}
\varepsilon_{\pi^*, \pi}^0 ({\bf k}) = \pm t \sqrt{ 3+ 2 \cos k_xa + 4 \cos \frac{k_x a}{2} \cos \frac{\sqrt{3} k_ya}{2}} - \mu,
\nonumber \\
\label{eq69}
\end{eqnarray}
and $t$ is the first neighbor hopping integral.

\begin{figure}[tb]
   \centerline{\includegraphics[width=15pc]{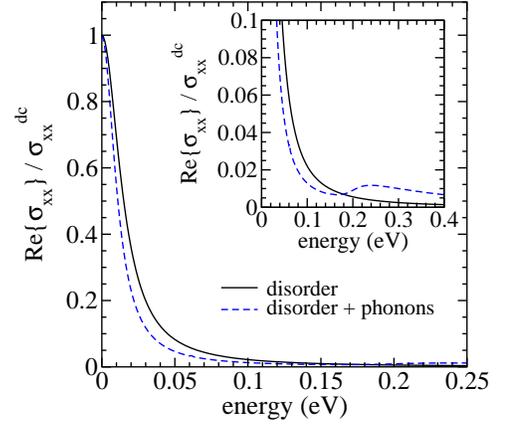}}
   \caption{(Color online) The real part of the intraband conductivity (\ref{eq54}) calculated beyond the Dirac cone approximation 
   for $E_{\rm F} = 0.5$ eV and $T = 50$ K.
   Solid line: the scattering by static disorder described by 
   $\hbar {\rm Im } \{ M_\alpha ({\bf k}_{\rm F}, \omega)\} \approx \hbar{\rm Im } \{  M_\alpha ({\bf k}_{\rm F}, 0) \} = \hbar \Gamma_1 = 15$ meV.
   Dashed line: the scattering by disorder and optical phonons, 
   $\hbar {\rm Im } \{ M_\alpha ({\bf k}_{\rm F}, \omega) \} = \hbar {\rm Im } \{ M^{\rm HE}_\alpha ({\bf k}_{\rm F},  \omega) \}
   + \hbar \tilde \Gamma_1$, with $|G_{op}|^2 = 0.5$ eV$^2$, $\hbar \omega_{op} = 0.2$ eV, 
   and $\hbar \tilde \Gamma_1 = 7.5$ meV, 
   [$\hbar{\rm Im } \{ M_\alpha ({\bf k}_{\rm F}, 0) \} = 15$ meV, again].
   The adiabatic parameter $\eta = \hbar \Sigma^i$ is taken to be $\eta = 20$ meV, and ${\bf k}_{\rm F} = (k_x,0)$, $k_x = 1.799$ \AA $^{-1}$.
   }
  \end{figure}

\subsection{Transverse conductivity sum rule}
The dynamical intraband conductivity ${\rm Re} \{\sigma_{\alpha \alpha} ({\bf q},\omega)\}$ from  Eq.~(\ref{eq54})
calculated in the approximation used in Fig.~3, $M_\alpha ({\bf k}_{\rm F}, \omega) \approx {\it i} \Gamma_1$,
is illustrated in Fig.~4 by the solid line.
	The integrated spectral weight is again in accordance with the partial transverse conductivity sum rule (\ref{eq20}).
	Namely, in this case
${\rm Re} \{ \sigma_{\alpha \alpha} ({\bf q},\omega) \}$ is nothing but
the sum of simple Lorentz functions ${\rm Im} \{ 1/(\omega + {\it i} \Gamma_1) \}$ multiplied by a function of ${\bf k}$
in which $n({\bf k}) \approx f({\bf k})$, and, consequently, integration over $\omega$ is trivial.

The dashed line represents ${\rm Re} \{\sigma_{\alpha \alpha} ({\bf q},\omega)\}$ in the case in which 
the scattering by acoustic phonons and by other electrons is taken aside. 
	The resulting memory function comprises two contributions $M_{\alpha}^\lambda ({\bf k},\omega)$,
where the index $\lambda = {\rm LE, HE}$ stands for the scattering by static disorder and by optical phonons, respectively.
	The phonon frequency is taken to be $\hbar \omega_{op} ({\bf q}) \approx \hbar \omega_{op} = 0.2$ eV and
the electron-phonon coupling function is $|G_{op}({\bf k},{\bf k}')|^2 \approx | G_{op}|^2 = 0.5$ eV$^2$, 
resulting in $\hbar {\rm Im} \{ M_\alpha ({\bf k}_{\rm F}, 0) \} = 15$ meV again.
	The integrated spectral weight is the same as in the first case.
	This characteristic of the dynamical conductivity is typical of the memory-function approaches.
	The $\omega$-dependent memory function $M_\alpha ({\bf k}_{\rm F}, \omega)$ leads to the redistribution of the (intraband) conductivity 
spectral weight over a wide energy range (up to 5 eV in Fig.~4).
	However, the integrated spectral weight is not changed.
	This is the first important conclusion regarding the memory-function $M_\alpha (\omega)$ from the generalized Drude formula.

\subsection{Hartree-Fock approximation}
In order to make the numerical calculations easier, the parameter $\eta$ in $M_\alpha^{[2]}({\bf k}_{\rm F}, \omega)$
from Fig.~4 is taken to be $\eta = 20$ meV.
	It is not hard to see that the physics behind such a parameter $\eta$ is simple.
	Namely, in the leading approximation, the recollection of higher-order contributions to $M_\alpha({\bf k}, \omega)$
corresponds to the replacement of the bare electron propagators in the diagrams $2A_1$, 
$2A_2$, and $2B$ on Fig.~2 by the renormalized propagators.
	In the spectral representation, this leads to the well-known Hartree-Fock approximation for $M_\alpha({\bf k}, \omega)$,
\begin{eqnarray}
&& \hspace{-12mm}
\hbar M_\alpha^{\rm H\hbox{-}F}({\bf k}, \omega) = \frac{1}{N} \sum_{\lambda {\bf k'}}  |G_\lambda({\bf k},{\bf k}')|^2 
\bigg(1 -\frac{v_\alpha ({\bf k}')}{v_\alpha ({\bf k})} \bigg)
\nonumber \\
&& \hspace{0mm}
\times  \int_{-\infty}^\infty \frac{d \varepsilon'}{2 \pi} \int_{-\infty}^\infty \frac{d \omega'}{2 \pi}  {\cal A} ({\bf k}', \varepsilon') 
{\cal B}_\lambda^0 ({\bf k}'-{\bf k},\omega')
\nonumber \\
&& \hspace{0mm}
\times  \frac{f^b (\omega') + f(\varepsilon')}{\hbar \omega  + {\it i} \eta  - \varepsilon' + \hbar \omega'}.
\label{eq70}
\end{eqnarray}
	Here, ${\cal B}_\lambda^0 ({\bf q}', \omega')$ is the bare phonon spectral function defined by
\begin{eqnarray}
&& \hspace{-8mm}
{\cal D}_\lambda^0 ({\bf q}', {\it i} \nu_m) = \int_{-\infty}^\infty \frac{d \omega'}{2 \pi}
\, \frac{{\cal B}_\lambda^0 ({\bf q}', \omega')}{{\it i} \nu_m-\omega'},
\label{eq71}
\end{eqnarray}
and ${\cal D}_\lambda^0 ({\bf q}', {\it i} \nu_m)$ is the bare phonon Green's function. 
	The next step in improving the expression (\ref{eq70}) might be simply to replace
${\cal D}_\lambda^0 ({\bf q}', {\it i} \nu_m)$ by the renormalized phonon propagator 
${\cal D}_\lambda ({\bf q}', {\it i} \nu_m)$.
	This is the $GW$ approximation for $M_\alpha({\bf k}, \omega)$.
	The comparison of Eq.~(\ref{eq55}), in which $\eta$ is replaced by the phenomenological parameter $\hbar \Sigma^i$,
with Eq.~(\ref{eq70}) shows that ${\rm Im} \{ \Sigma^{\rm H\hbox{-}F}({\bf k}, \omega) \} \approx -\Sigma^i$, i.e.,
$1/\Sigma^i$ can be understood as the phenomenological electron lifetime from ${\cal A} ({\bf k}, \varepsilon)$.

\begin{figure}[tb]
   \centerline{\includegraphics[width=17pc]{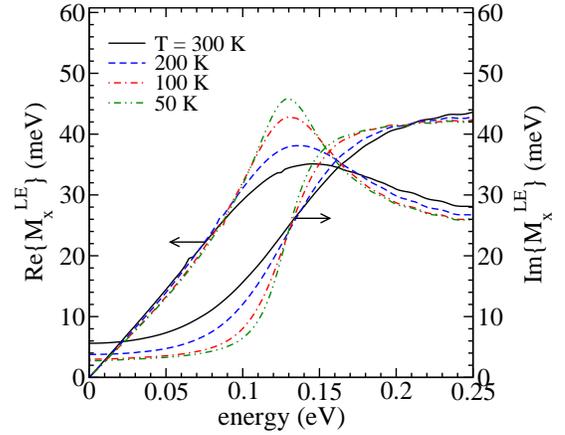}}
   \caption{(Color online) The real and imaginary parts of the memory function 
   $M^{{\rm LE}[2]}_\alpha ({\bf k}_{\rm F},  \omega)$ as a function of temperature,
   for ${\bf k}_{\rm F} = (k_x,0)$, $k_x = 1.799$ \AA $^{-1}$,
   $t=2.52$ eV,  $E_{\rm F} = 0.5$ eV, $|C_{ac}|^2 = 0.025$ eV$^2$, 
   $\hbar \omega_{ac} = 30$ meV, and $\eta = 10$ meV.
   }
  \end{figure}

\begin{figure}[tb]
   \centerline{\includegraphics[width=17pc]{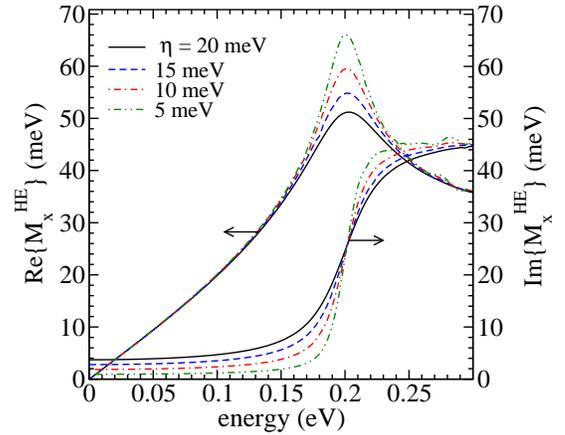}}
   \caption{(Color online) 
   The real and imaginary parts of the memory function 
   $M^{{\rm HE}[2]}_\alpha ({\bf k}_{\rm F},  \omega)$ as a function of the parameter $\eta = \hbar \Sigma^i$,
   for ${\bf k}_{\rm F} = (k_x,0)$, $k_x = 1.799$ \AA $^{-1}$,
   $t=2.52$ eV,  $E_{\rm F} = 0.5$ eV, $|G_{op}|^2 = 0.25$ eV$^2$, 
   $\hbar \omega_{op} = 0.2$ eV, and $T=50$ K.
   }
  \end{figure}

\subsection{$\omega$-dependent effective mass}
The most common form of the generalized Drude formula used in weakly interacting isotropic systems is the following \cite{Basov11,Uchida91}
\begin{eqnarray}
&& \hspace{-10mm}
\sigma (\omega) 
= \frac{{\it i} e^2}{m(\omega)} \frac{n}{\omega  + {\it i}/\tau(\omega)}.
\label{eq72}
\end{eqnarray}
	Here, $n$ is the nominal concentration of conduction electrons/holes, $\tau(\omega)$ is the $\omega$-dependent relaxation time, 
and $m(\omega) = m(1 + \lambda(\omega))$ is the $\omega$-dependent effective mass.
	In weakly interacting anisotropic systems, it can be rewritten in the form
\begin{eqnarray}
&& \hspace{-10mm}
\sigma_{\alpha \alpha} (\omega) 
= \frac{{\it i} e^2}{m_{\alpha \alpha}(\omega)} \frac{n_{\alpha \alpha}^{\rm intra}}{\omega  + {\it i}/\tau_\alpha(\omega)},
\label{eq73}
\end{eqnarray}
with $m_{\alpha \alpha}(\omega) = m(1 + \lambda_\alpha(\omega))$.
	The comparison with Eq.~(\ref{eq54}) shows that $\lambda_\alpha(\omega) \approx \lambda_\alpha ({\bf k}_{\rm F},\omega)$.
	Here, $\lambda_\alpha ({\bf k},\omega) =  {\rm Re} \{M_\alpha ({\bf k}, \omega) \}/\omega$ 
is the usual notation for the dimensionless electron-phonon coupling constant.

In order to illustrate the dependence of $\lambda_\alpha(\omega)$ in heavily doped graphene on the model parameters, we show in Figs.~5 and 6 
the real and imaginary parts of $M_{\alpha}^{\rm LE} ({\bf k}_{\rm F},\omega)$ (scattering by acoustic phonons) and $M_{\alpha}^{\rm HE} ({\bf k}_{\rm F},\omega)$
(scattering by optical phonons) for typical values of the model parameters.
	The phonon dispersions are approximated by  $\hbar \omega_{ac} ({\bf q}) \approx \hbar \omega_{ac}qa$, 
$\hbar \omega_{op} ({\bf q}) \approx \hbar \omega_{op}$ and the electron-phonon coupling functions are assumed to be 
$G_{ac}({\bf k},{\bf k}_+) \approx C_{ac} qa$, $G_{op}({\bf k},{\bf k}_+) \approx G_{op}$.
	For the values of the parameters used in Figs,~5 and 6, we obtain $\lambda = \lambda_\alpha ({\bf k}_{\rm F},0) \approx 0.5$,
resulting in $m(0) = 1.5 m$ (the same value of $\lambda$ is obtained for the case shown in Fig.~4).
	These two figures show that $\lambda = \lambda^{\rm LE} + \lambda^{\rm HE}$ is largely unaffected by
both temperature and the damping energy $\eta = \hbar \Sigma^i$.

\section{Lightly doped graphene}
Lightly doped graphene is an interesting example of multiband electronic systems in which the ratio between the threshold energy for interband electron-hole excitations 
($2E_{\rm F}$) and the energy of optical phonons [$\hbar \omega_{op} ({\bf q}) \approx \hbar \omega_{op}$] 
can be easily tuned by the electric field effect \cite{Novoselov05,Zhang05,Li08}.
	For the scattering by phonon modes,
we can introduce the interband memory functions $M_{\alpha}^{LL'} ({\bf k},\omega)$, $ L \neq L'$, by using the procedure illustrated in Sec.~VI\,B.
	These functions are expected to have the $\omega$-dependence similar to the $\omega$-dependence of the intraband memory functions $M_{\alpha}^{LL} ({\bf k},\omega)$.
	According to Figs.~5 and 6, this means that in the interband relaxation-time approximation, ${\rm Im} \{ M_{\alpha}^{LL'} ({\bf k},\omega) \} \approx \Gamma_2$,
there will be two different regimes depending upon whether
$2E_{\rm F} > \hbar \omega_{op}$ ($\Gamma_2 \gg \Gamma_1$ in this case) or $2E_{\rm F} \ll \hbar \omega_{op}$ (where $\Gamma_2 \approx \Gamma_1$).
	However, to better understand
the relaxation processes in the interband channel, we have to include in the self-consistent RPA equation (\ref{eq30}) the scattering by other electrons as well.
	The detailed discussion of this question will be given in a separate presentation \cite{KupcicUP}.

\begin{figure}[tb]
   \centerline{\includegraphics[width=15pc]{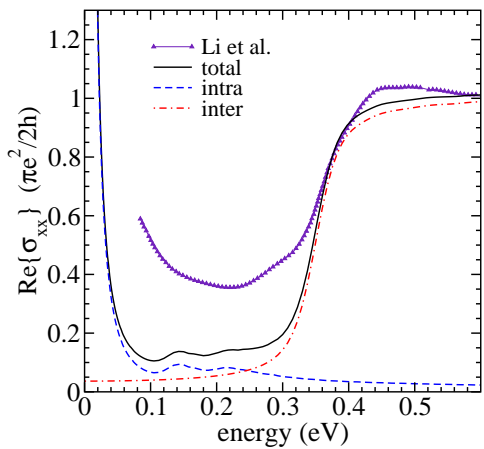}}
   \caption{(Color online) 
   The real part of the two-band dynamical conductivity in lightly doped graphene,
   for $E_{\rm F} = 0.175$ eV and $T = 50$ K.
   The scattering by static disorder is described by $\hbar \tilde \Gamma_1 = 1.6$ meV, the scattering by acoustic and optical phonons by 
   $\hbar {\rm Im } \{ M^{\rm LE}_\alpha ({\bf k}_{\rm F},  \omega) \}$ and
   $\hbar {\rm Im } \{ M^{\rm HE}_\alpha ({\bf k}_{\rm F},  \omega) \}$, and $\hbar \Gamma_2 = 20 $ meV is the interband relaxation rate.
   The parameters in the memory function are the same as in Figs.~5 and 6, with ${\bf k}_{\rm F} = (k_x,0)$, $k_x = 1.736$ \AA $^{-1}$.
   Experimental results, taken at $T = 40$ K, are from Ref.~\cite{Li08}.
   }
  \end{figure}

Here, we are focused on the interband relaxation-time approximation, for $\Gamma_1 \ll \Gamma_2$.
	Figure 7 shows the two-band dynamical conductivity in such a case ($E_{\rm F} = 0.175$ eV).
	The scattering by other electrons is taken aside and the relaxation processes in the interband channel are treated in the relaxation-time approximation.
	The scattering by acoustic and optical phonons is described in the same way as in Figs.~5 and 6 (with $\eta = 10$ meV).

This figure shows that if we are interested in low-energy electrodynamic properties of conduction electrons in systems 
in which the threshold energy for interband electron-hole excitations is of the order of the optical phonon frequencies 
(or other high-frequency boson modes), we have to subtract from experimental spectra both the interband contributions and 
the high-energy part of the intraband memory function.
	The resulting low-energy part of $\sigma_{\alpha \alpha} ({\bf q},\omega)$ can be shown in the form
\begin{eqnarray}
&& \hspace{-10mm}
\sigma_{\alpha \alpha}^{\rm LE} ({\bf q},\omega) \approx 
\frac{{\it i} e^2}{m}\frac{1}{V} \sum_{{\bf k} \sigma} m v_\alpha^2 ({\bf k})\bigg(-\frac{\partial n({\bf k})}{\partial \varepsilon_0({\bf k})}\bigg)
\nonumber \\
&& \hspace{8mm}
\times
\frac{1}{\alpha^{\rm HE}({\bf k})}\frac{1}{\omega  +   M_\alpha^{\rm LE}({\bf k},\omega)/\alpha^{\rm HE}({\bf k})}, 
\label{eq74}
\end{eqnarray}
where $\alpha^{\rm HE}({\bf k}) = 1+\lambda_\alpha^{\rm HE}({\bf k},0)$.

The same form of $\sigma_{\alpha \alpha}^{\rm LE} ({\bf q},\omega)$ is expected for coherent low-energy conductivity 
of various low-dimensional strongly correlated electronic systems, such as the cuprate superconductors \cite{Uchida91,Mirzaei13,Schrieffer07}.
	In all such cases, we can perform the memory-function analysis based on the function (\ref{eq74}), estimate both
$\alpha^{\rm HE}({\bf k})$ and the frequency dependence of the real and imaginary parts of $M_\alpha^{\rm LE}({\bf k},\omega)$
[or their averages over the Fermi surface $\alpha^{\rm HE}$ and $M_\alpha^{\rm LE}(\omega)$],
and identify the most intense low-energy scattering channels.
	Such an analysis must be completed with temperature measurements of the transport coefficients, in the first place, the dc
resistivity $\rho^{\rm dc} \approx 1/\sigma_{\alpha \alpha}^{\rm LE} ({\bf q},\omega\!=\!0)$.
	Therefore, to understand the low-energy physics in such systems, we have to examine carefully all scattering channels in the 
intraband and interband memory functions.

\section{Conclusion}
In this article, we have presented the generalization of the common self-consistent RPA equation to re-derive the memory-function conductivity formula 
in a general weakly interacting multiband electronic system.
	The generalized RPA equations are shown to be integral equations which can be easily solved by iteration.
	The resulting conductivity formula and the structure of ${\bf k}$- and $\omega$-dependent intraband memory function 
turn out to be the same as that obtained by using a more general approach based and the quantum kinetic equations.

The results are applied to heavily doped and lightly doped graphene.
	It is shown that the scattering of conduction electrons by phonons leads to the redistribution of 
the intraband spectral weight over a wide energy range, in a way consistent with the partial transverse conductivity sum rule.
	It is also shown that the present form of the intraband memory function includes  the scattering by quantum fluctuations 
of the lattice, at variance with the standard semiclassical expressions for the intraband relaxation rate, 
where this scattering channel is absent.
	Finally, it is illustrated how the effective generalized Drude formula can be used to study low-energy dynamical conductivity 
in multiband electronic systems in which the threshold energy for the interband electron-hole excitations is of the order 
of the optical phonon energies.
	This simplified conductivity formula is expected to be of importance in analyzing coherent contributions 
to the intraband conductivity in different strongly correlated electronic systems, such as the cuprate superconductors.

\section*{Acknowledgments}
This research was supported by the Croatian Ministry of Science, Education and Sports
under Project No. 119-1191458-0512 and the University of Zagreb Grant No. 20281315

\appendix

\section{Common memory-function approach}
In order to re-derive the generalized Drude conductivity formula from Ref.~\cite{Gotze72},
we consider the microscopic real-time RPA irreducible current-current correlation function $\pi_{\alpha \alpha}({\bf q},t)$
from Eq.~(\ref{eq13}).
	The integration by parts with respect to time of the Fourier transform $\pi_{\alpha \alpha}({\bf q},\omega)$ gives 
\cite{Kupcic14}
\begin{eqnarray}
&& \hspace{-11mm}
\pi_{\alpha \alpha}({\bf q},\omega) = -\frac{1}{(\hbar \omega)^2} \frac{1}{V} \big[ \Phi_{\alpha \alpha}(\omega) 
- \Phi_{\alpha \alpha}(0) \big],
\label{eqA1}
\end{eqnarray}
where
\begin{eqnarray}
&& \hspace{-11mm}
\Phi_{\alpha \alpha}(\omega) 
= \langle \langle [\hat J_\alpha ({\bf q}),H] ; [\hat J_\alpha (-{\bf q}),H] \rangle \rangle_\omega^{\rm irred}.
\label{eqA2}
\end{eqnarray}
	After inserting
\begin{eqnarray}
&& \hspace{-10mm}
[\hat J_\alpha ({\bf q}),H] \approx [\hat J_\alpha ({\bf q}),H'] =
[\hat J_\alpha ({\bf q}), H'_1 + H_2']
\label{eqA3}
\end{eqnarray}
in Eq.~(\ref{eqA2}), we obtain the $(H')^2$ contribution to the dynamical conductivity $\sigma_{\alpha \alpha}({\bf q},\omega)$.
	It is given by the  Kubo formula (\ref{eq17})
in which the current-current correlation function $\pi_{\alpha \alpha}({\bf q},\omega)$ is replaced by its high-energy contribution
\begin{equation}
\pi_{\alpha \alpha}^{(2)}({\bf q},\omega) = -\frac{1}{(\hbar \omega)^2} \frac{1}{V}
\big[ \Phi^{(2)}_{\alpha \alpha}(\omega) - \Phi^{(2)}_{\alpha \alpha}(0) \big],
\label{eqA4}
\end{equation}
with
\begin{equation}
\Phi^{(2)}_{\alpha \alpha}(\omega) = \langle \langle [\hat J_\alpha ({\bf q}),H'] ; 
[\hat J_\alpha (-{\bf q}),H'] \rangle \rangle_\omega^{\rm irred}.
\label{eqA5}
\end{equation}
	Therefore, the common memory-function approach leads to the generalized Drude formula from the main text,
Eq.~(\ref{eq67}),
in which $M_{\alpha} (\omega)$ is given by
\begin{eqnarray}
&& \hspace{-12mm}
\hbar M_{\alpha} (\omega) = \frac{1}{n^{\rm intra}_{\alpha \alpha}}
\frac{m}{e^2 \hbar \omega} \frac{1}{V} \big[ \Phi^{(2)}_{\alpha \alpha}(\omega) - \Phi^{(2)}_{\alpha \alpha}(0) \big].
\label{eqA6}
\end{eqnarray}

For example, for $H'=H'_{1a}$, a straightforward calculation gives
\cite{Gotze72}
\begin{eqnarray}
&& \hspace{-7mm}
\hbar M^{[2]}_{\alpha} (\omega) = \frac{1}{n^{\rm intra}_{\alpha \alpha}}
\frac{1}{V} \sum_{{\bf k} \sigma} m v_\alpha^2 ({\bf k})
(-)\frac{1}{N} \sum_{\lambda {\bf k'}}\bigg(1 -\frac{v_\alpha ({\bf k}')}{v_\alpha ({\bf k})} \bigg)^2
\nonumber \\
&& \hspace{5mm}
\times |G_\lambda({\bf k},{\bf k}')|^2 \frac{1}{\varepsilon_0({\bf k},{\bf k}') + \hbar \omega_{\lambda {\bf q}'}}
\nonumber \\
&& \hspace{5mm}
\times \sum_{s = \pm 1} 
\frac{s[(1+f^b)(1-f)f' - f^b (1-f')f] }{
s(\hbar \omega + {\it i} \eta) + \varepsilon_0({\bf k},{\bf k}') + \hbar \omega_{\lambda {\bf q}'}},
\label{eqA7}
\end{eqnarray}
where $f = f({\bf k})$,   $f' = f({\bf k}')$, and  $f^b =f^b (\omega_{\lambda {\bf q}'})$.
	After using the microscopic reversibility principle, this expression transforms into
\begin{eqnarray}
&& \hspace{-10mm}
\hbar M^{[2]}_{\alpha} (\omega) = \frac{1}{n^{\rm intra}_{\alpha \alpha}}
\frac{1}{V} \sum_{{\bf k} \sigma} m v_\alpha^2 ({\bf k})
(-)\frac{1}{N} \sum_{\lambda {\bf k'}} \beta (1-f')f
\nonumber \\
&& \hspace{5mm}
\times \bigg(1 -\frac{v_\alpha ({\bf k}')}{v_\alpha ({\bf k})} \bigg)^2 |G_\lambda({\bf k},{\bf k}')|^2 
\nonumber \\
&& \hspace{5mm}
\times \sum_{s = \pm 1} 
\frac{s f^b(\omega_{\lambda {\bf q}'})}{
s(\hbar \omega + {\it i} \eta) + \varepsilon_0({\bf k},{\bf k}') + \hbar \omega_{\lambda {\bf q}'}}.
\label{eqA8}
\end{eqnarray}

Similarly, for $H'=H'_2$, we obtain
\begin{eqnarray}
&& \hspace{-5mm}
\hbar M^{[4]}_{\alpha} (\omega) = \frac{1}{n_{\alpha \alpha}^{\rm intra} }
\frac{1}{V} \sum_{{\bf k} \sigma} 
m v_\alpha^2 ({\bf k}) \sum_{{\bf k'}{\bf q}} \frac{|\varphi({\bf q})|^2}{V^2} 
\nonumber \\
&& \hspace{10mm}
\times \frac{1}{v_\alpha({\bf k})}\big[v_\alpha({\bf k}) + v_\alpha({\bf k}'_+) - v_\alpha({\bf k}')
-v_\alpha({\bf k}_+)  \big]
\nonumber \\
&& \hspace{5mm}
\times \sum_{s = \pm 1} \frac{2s}{\varepsilon_0({\bf k},{\bf k}') + \varepsilon_0({\bf k}'_+,{\bf k}_+)}
\nonumber \\
&& \hspace{0mm}
\times \frac{(1-f_+)(1-f')f_+'f - (1-f)(1-f_+')f'f_+}{
\hbar \omega + {\it i} \eta + s \varepsilon_0({\bf k},{\bf k}') + s \varepsilon_0({\bf k}'_+,{\bf k}_+)},
\label{eqA9}
\end{eqnarray}
where $f_+ = f({\bf k}_+)$ and $f'_+ = f({\bf k}'_+)$.

It is easily seen that Eq.~(\ref{eqA8}) is directly related to the result of 
the variational approach for the relaxation rate $\hbar/\tau_{\rm tr}$ \cite{Gotze72,Ziman72}:
\begin{eqnarray}
&& \hspace{-5mm}
\frac{\hbar}{\tau_{\rm tr}} = \frac{1}{n^{\rm intra}_{\alpha \alpha}}
\frac{1}{V} \sum_{{\bf k} \sigma} m v_\alpha^2 ({\bf k})
\frac{1}{N} \sum_{\lambda {\bf k'}} \beta (1-f')f
\bigg(1 -\frac{v_\alpha ({\bf k}')}{v_\alpha ({\bf k})} \bigg)^2 
\nonumber \\
&& \hspace{5mm}
\times  |G_\lambda({\bf k},{\bf k}')|^2  f^b (\omega_{\nu {\bf q}'}) 
2 \pi \delta(\varepsilon_0({\bf k},{\bf k}') + \hbar \omega_{\lambda {\bf q}'}).
\label{eqA10}
\end{eqnarray}
	After using the thematic simplification $ \beta (1-f')f \approx -\partial f({\bf k})/\partial \varepsilon_0({\bf k})$
in Eq.~(\ref{eqA8}) and the relation (\ref{eq66}), we obtain the following expression for the ${\bf k}$-dependent memory function:
\begin{eqnarray}
&& \hspace{-5mm}
\hbar M^{[2]}_{\alpha} ({\bf k}, \omega) =  -\frac{1}{N} \sum_{\lambda {\bf k'}} 
|G_\lambda({\bf k},{\bf k}')|^2 \bigg(1 -\frac{v_\alpha ({\bf k}')}{v_\alpha ({\bf k})} \bigg) 
\nonumber \\
&& \hspace{5mm}
\times \sum_{s = \pm 1} \sum_{s' = \pm 1} \frac{f^b (\omega_{\lambda {\bf k}-{\bf k}'})}{
\hbar \omega + {\it i} \eta + s \varepsilon_0({\bf k},{\bf k}') + s' \hbar \omega_{\lambda {\bf k}-{\bf k}'}}.
\nonumber \\
\label{eqA11}
\end{eqnarray}
	The term $ f(s s' \varepsilon_0({\bf k}'))$ from Eq.~(\ref{eq55}) is missing in both of these two standard textbook expressions.

Another disadvantage of the common memory-function approach is that the memory function (\ref{eqA6}) is second order in perturbation $H'$.
	Consequently, to study the phenomena such as the SDW instability \cite{Schwartz98} 
or the BCS instability \cite{Mirzaei13}
of the electronic subsystem, the scattering by soft phonons \cite{Degiorgi91}, or by intraband plasmon modes, we have to go beyond this approximation.
	It is thus necessary to develop high-order perturbation theory 
for the electron-hole self-energy which recollects the most singular contributions 
in a systematic way.	
	The Green's function method presented in Sec.~VI represents one possible way to do this.

\end{document}